\documentclass{iopart}
\usepackage{graphicx}
\usepackage{pfnote}
\usepackage{amsfonts}
\usepackage{amssymb}
\usepackage{bm,bbm}
\usepackage{euscript}
\usepackage{hyperref}

\def \ltsim    {\:\relax{\mathrel{\mathpalette\oversim <}}\:}
\def\oversim#1#2{\lower4pt\vbox{\baselineskip0pt \lineskip1.5pt
            \ialign{$\mathsurround=0pt#1\hfil##\hfil$\crcr#2\crcr\sim\crcr}}}

\def\1{{{\mathbbm 1}}}
\def\6{\langle}
\def\9{\rangle}

\def\half{\mbox{$1\over2$}}

\def\bnab{{\mbox{\boldmath{$\nabla$}}}}
\def\anab{{\mbox{\sf\boldmath{{$\nabla$}}}}}

\def\beq{\begin{equation}}
\def\eeq{\end{equation}}

\def\bal{\mbox{\boldmath $\alpha$}}
\def\bep{\mbox{\boldmath $\epsilon$}}

\def\sg{\mbox{\sl g}}

\def\bof{\mathbf{f}}
\def\bk{{\bf k}}
\def\bp{{\bf p}}

\def\bz{{\bf z}}

\def\hbb{{\hat{\mathbf{b}}}}
\def\hbk{\hat{\bk}}
\def\hbp{\hat{\bp}}

\def\hbz{\hat{\bz}}
\def\hbf{{\hat{\bof}}}
\def\hbk{{\hat{\bk}}}

\def\bE{\mathbf{E}}

\def\eR{\EuScript{R}}

\newcommand{\Omegab}{\mbox{\boldmath$\Omega$}}
\newcommand{\omegab}{\mbox{\boldmath$\omega$}}

\begin{document}
\renewcommand{\thefootnote}{\arabic{footnote}}

\title[Fundamental quantum optics experiments conceivable with satellites]{Fundamental quantum optics experiments conceivable with satellites --- reaching relativistic distances and velocities}

\author{David~Rideout$^{1,2,3\,\ast}$, Thomas~Jennewein$^{2,4\,\dagger}$, Giovanni~Amelino-Camelia$^5$, Tommaso~F Demarie$^6$, Brendon~L~Higgins$^{2,4}$, Achim~Kempf$^{\,2,3,7}$, Adrian~Kent$^{3,8}$,
  Raymond~Laflamme$^{2,3,4}$, Xian~Ma$^{2,4}$, Robert~B~Mann$^{2,4}$,
  Eduardo~Mart\'in-Mart\'inez$^{2,4}$, Nicolas~C~Menicucci$^{3,9}$, John~Moffat$^{3}$, Christoph~Simon$^{10}$, Rafael~Sorkin$^{3}$,
  Lee~Smolin$^3$, Daniel~R~Terno$^{6}$} 

\address{$^1$ current address: Department of Mathematics,
University of California / San Diego,
La Jolla, CA, USA\\
$^\ast$ E-mail: drideout@math.ucsd.edu}

\address{$^2$ Institute for Quantum Computing,
 Waterloo, ON, Canada}
\address{$^\dagger$ E-mail: thomas.jennewein@uwaterloo.ca}

\address{$^3$ Perimeter Institute for Theoretical Physics,
  Waterloo, ON, Canada}

\address{$^4$ Department of Physics and Astronomy,
University of Waterloo,
 Waterloo, ON, Canada}

\address{$^5$ Departimento di Fisica,
Universit\`a di Roma ``La Sapienza'',
 Rome, Italy}

\address{$^6$ Department of Physics and Astronomy, Macquarie University, Sydney, NSW, Australia}

\address{$^7$ Department of Applied Mathematics and Department of Physics and Astronomy,
University of Waterloo,
 Waterloo, ON, Canada}

\address{$^8$ Centre for Quantum Information and Foundations, DAMTP,
University of Cambridge, Cambridge, U.K.}

\address{$^9$ School of Physics, The University of Sydney, NSW, Australia}

\address{$^{10}$ Institute for Quantum Information Science and Department of
  Physics and Astronomy, University of Calgary, Calgary, AB, Canada}

\begin{abstract}
  Physical theories are developed to describe phenomena in particular
  regimes, and generally are valid only within a limited range of scales.
  For example, general relativity provides an effective description of the
  Universe at large length scales, and has been tested from the cosmic scale
  down to distances as small as 10 meters~\cite{Atomin,Atomin2}.  In
  contrast, quantum theory provides an effective description of physics at
  small length scales.  Direct tests of quantum theory have been performed at
  the smallest probeable scales at the Large Hadron Collider, ${\sim}
  10^{-20}$ meters, up to that of hundreds of
  kilometers~\cite{UTSWSLBJPTOFMRSBWZ07}.  Yet, such tests fall short of the
  scales required to investigate potentially significant physics that arises
  at the intersection of quantum and relativistic regimes.  We propose to
  push direct tests of quantum theory to larger and larger length scales,
  approaching that of the radius of curvature of spacetime, where we begin to
  probe the interaction between gravity and quantum phenomena.  In
  particular, we review a wide variety of potential tests of fundamental
  physics that are conceivable with artificial satellites in Earth orbit and
  elsewhere in the solar system, and attempt to sketch the magnitudes of
  potentially observable effects.  The tests have the potential to determine
  the applicability of quantum theory at larger length scales, eliminate
  various alternative physical theories, and place bounds on phenomenological
  models motivated by ideas about spacetime microstructure from quantum
  gravity.  From a more pragmatic perspective, as quantum communication
  technologies such as quantum key distribution advance into Space towards
  large distances, some of the fundamental physical effects discussed here
  may need to be taken into account to make such schemes viable.

\end{abstract}

\date{\today}
\maketitle

\makeatletter\section*{\contentsname}    \@starttoc{toc}\makeatother
\pagestyle{headings}

\section{Introduction}
\label{timeframes.sec}

Our knowledge is ultimately restricted by the
boundaries of what we have explored by direct observation or experiment.
Experiments conducted within
previously inaccessible regimes have often revealed new aspects of the
Universe, facilitating new insights into its fundamental operation.  Examples
of this pervade the history of physical science. Recently, the theories
of general relativity and quantum
field theory have arisen to describe aspects of
the Universe that can only be accessed experimentally within the regimes of the
very large and the very small, respectively.
The spectacular array of useful technologies brought about owing to the formulation of these theories over the previous century are vociferous testament to the utility of expanding our experimental horizons.

The success of these two theories also confronts us with a formidable challenge.
On one hand, quantum theory
excellently describes the behaviour of physical systems at small length
scales.
On the other hand, general relativity theory excellently describes
systems involving very large scales: long distances, high accelerations, and
massive bodies.\footnote{At the very smallest of physical scales --- the Planck scale --- one expects the
  gravitational interaction to become comparable to all others, so that
  quantum and gravitational effects are both simultaneously manifest.  The same may also
  occur at the largest physical scale, that of the cosmos as a whole, wherein
  quantum effects may account for formation of large scale structures
  and the cosmic acceleration~\cite{SorkinLambda}.}
Each of these theories has successfully weathered copious experimental tests
independently, yet the two theories are famously incompatible in their
fundamental assertions.
One expects that both theories are 
limiting cases of one set of overarching laws of physics. However,
the tremendous experimental success of quantum theory and general relativity,
i.e., their enormous individual ranges of validity, 
makes it extremely difficult to find
experimental evidence that points us towards such
unifying laws of physics --- a fully quantum theory of
gravity.

Theoretical research 
has yielded intriguing indications about this sought-after unifying theory of
quantum gravity. For example, studies of quantum effects in the presence of
black holes, such as Hawking radiation and 
evaporation,  indicate that it will be crucial to understand how the flow and transformations of information are impacted by relativistic and quantum effects, with the notion of entanglement playing a central role.

Ultimately, however, the field of quantum gravity will require more
experimental guidance.
So far, the best potential for solid experimental
evidence for full-blown quantum gravitational effects 
stems from observations
of the cosmic microwave background (CMB). 
However, while of the highest interest, the observation of
quantum gravity effects in the CMB would still constitute only a passive
one-shot experimental observational opportunity --- we cannot repeat the big
bang.

In this paper we thus envisage possible
avenues for active experimental probes of quantum phenomena at large length
scales, towards those at which gravitational effects will play an
increasingly significant role.
The aim in the short term is to probe more-or-less solid theoretical
expectations, while in the longer term to explore physical regimes in which
the predictions of theory are not as clear.

To begin making inroads, it seems necessary to test the behaviour of quantum systems, particularly those with entanglement, while these systems possess high speeds and are separated by large distances. On Earth, tests of quantum entanglement have been performed at distances up to 144\,km~\cite{UTSWSLBJPTOFMRSBWZ07}. While this is a significant achievement, it falls short of the large scale relevant for relativistic considerations. Additionally, these tests were performed with stationary detectors. While it is difficult to perform tests with moving detectors, it is conceivable to measure entanglement with beamsplitters moving at up to 1000\,m/s (${\sim}10^{-6} c$)~\cite{PhysRevLett.88.120404}. However, this also falls short --- one would need to perform entanglement tests in which the detectors are in relative motion at speeds nearer lightspeed ($c$) where relativistic effects become significant.
It is interesting to note that
active laboratory measurements of gravity at small scales using atomic interferometers
have also been proposed~\cite{Atomin}.

\begin{figure}[!phtb]
\begin{center}
\includegraphics[width=0.75\textwidth]{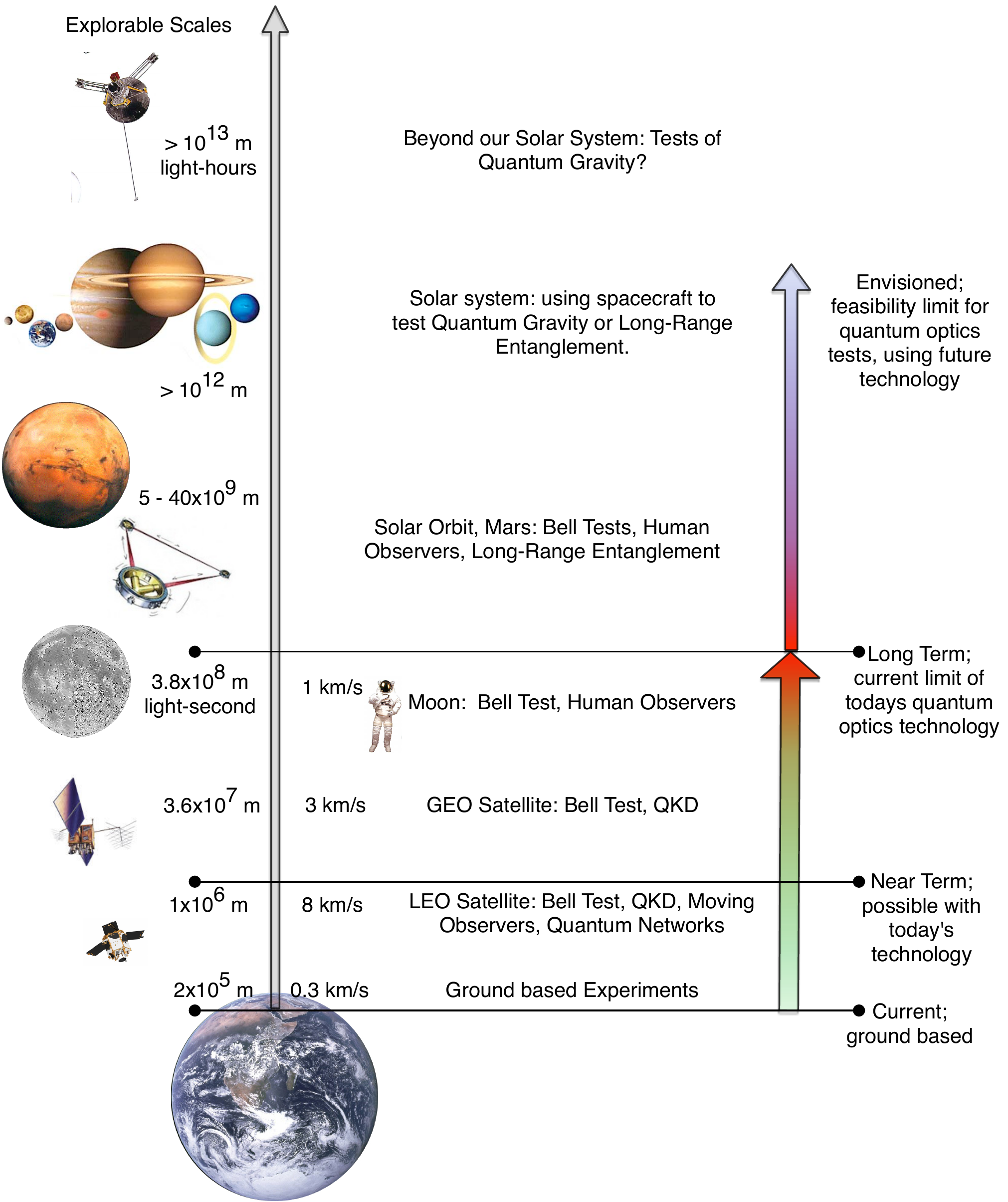}
\caption{Overview of the distance and velocity scales achievable in a space
  environment explorable with man-made systems, with some possible quantum optics experiments at each given distance.}
\label{overviewScales}
\end{center}
\end{figure}

Quantum repeater networks are a promising candidate for the long distance
dissemination of quantum entanglement~\cite{Sangouard2011}.
However, the study of quantum repeaters shows that even with optimistic
estimates, reaching 1000\,km will be a huge challenge.  Even if quantum transmission technology is developed that is capable of transmitting entangled
systems around the Earth, the maximum separation between detectors that can
be achieved 
is bounded by the Earth's diameter: around 13,000\,km.

To achieve tests at greater distances and speeds, one needs to move off of planet Earth, and into Space.  It is conceivable that in the not-too-distant future one could perform quantum entanglement tests at the scale of inter-planetary distances, with the associated velocities. For the nearer term, the next step is to perform quantum experiments that utilize Earth-orbiting satellite platforms. A satellite in low Earth orbit (LEO), for example, will allow distances greater than $10^6$\,m and relative speeds of two detectors of ${\sim} 10^{-5}c$.

Satellite missions for quantum communications have been considered in various configurations~\cite{nordholt:116, Armengol2008, RTGK02, Xin20052011, Toyoshima:2011lr}, and some scientific tests utilizing such satellites have been proposed. The Canadian Space Agency (CSA) and the Institute for Quantum Computing (IQC) have been participating in ongoing studies, dubbed QEYSSAT~\cite{jennewein2012}, emphasizing the wider and long term context of such missions for science as well as quantum applications such as global-scale quantum key distribution (QKD).

These and other proposals for satellite-based quantum apparatuses open a
door to distances and velocities that are either prohibitively impractical
or simply impossible to achieve on the ground. Here we describe a number of
ideas stemming from a series of discussions that took place at the
Perimeter Institute for Theoretical Physics, which focused on what tests of
fundamental quantum physics could be achieved
with such experimental setups.  We consider a variety of scenarios,
illustrated in Figure~\ref{overviewScales}, of which some will be accessible
with today's technologies, such as a single satellite at LEO altitudes (500--1000\,km).  Experiments at larger
distances 
will be possible only on a longer time frame, owing to their complexity and
the advanced technologies that are required, and include systems in 
geostationary (GEO) orbits (36,000\,km) or even Earth-Moon distances 
(380,000\,km). Visionary experiments involving distances at the scale of the
Earth's distance to the Sun (150\,Gm) are conceivable, but require technologies yet to be developed.

Here we consider experimental scenarios that are visionary in nature, focusing on the scientific novelty of such experiments and their capacity to offer significant new insights, and intentionally avoiding dwelling on technological or financial limitations. It follows that these scenarios may be achievable within varying timeframes. The space science community may at some point consider these proposals within the context of future missions.
Furthermore, we attempt to avoid 
bias with regard to expectations of experimental outcomes
by also considering several ideas which are based on 
unconventional physical theories.

For each proposed experiment we explain the basic idea and physical concepts that it
probes, including references for further details.  We also attempt to provide
some characterization of the magnitude of the expected effect.
We wish to note that this  paper resembles a review article, as it covers a wide spectrum of
physical effects, contributed by multiple authors from a number of
perspectives.
In order to give a better indication of the primary contributor for each section, we  indicate the primary authors at the end of the paper.

We hope to encourage physicists to take one or more of these experiment
concepts and work towards establishing the details of how it may work in
practice, and more carefully examine the magnitude of observable effects that
may be observed.  Similar questions have already been posed for the Space-QUEST project, which aims to place an entangled photon source on the International Space Station~\cite{KAJBPLZ03,Ursin_EPN_Space-QUEST}. Our analysis goes beyond that work and studies various possible science experiments on a broader scope.

The paper is organized as follows.  In Section~\ref{sec.classification} we
define a broad classification scheme for the proposed experiments.  In
Section~\ref{sec.summary} we present a table summarizing the list of
experiments, including some indication of the practical feasibility
of each.
The experiments themselves are organized into broad categories based upon the
nature of the physical theories which the experiment is designed to probe.
In Section~\ref{sec.entanglement} we discuss EPRB-type tests of the Bell
inequalities.  In Section~\ref{sec.entanglementSRGR} we consider the effect
of both special
and general relativity.
In Section~\ref{sec.qft} we discuss tests of quantum field theory in
accelerated frames.
Section~\ref{sec:q_grav_expts} considers possible effects motivated by
quantum gravity.
In Section~\ref{sec.qcommunication} we include some experiments whose
motivations are directed towards developing useful quantum communication
technologies, such as quantum key distribution and quantum teleportation.
Section~\ref{sec.spekkens} considers some variants on the usual EPRB-type
setup which may be useful to consider for a number of the Bell-test
scenarios.
Finally, in Section~\ref{sec.technology} we provide some technical details on the
current state of the art for quantum optics experiments in Space, and make
some concluding remarks in Section~\ref{sec.conclusion}.

\subsection{Classification of experiments}\label{sec.classification}

In order to give the reader some initial indicator of the nature of a proposed experiment,
 we introduce a 
broad classification scheme
that roughly characterizes the feasibility and ambitiousness of each test.  The classes are defined as follows:
\begin{itemize}
 \item Level-1 experiments verify well-established physics in a new regime, often at larger length scales than have yet been probed experimentally.

 \item Level-2 experiments test physics that is somewhat less certain.
An
example is an experiment whose predicted outcome involves the parallel transport of
spins in curved spacetime.  
It seems pretty clear along which spacetime path
 the spins should be transported, however physical phenomena whose
outcomes involve such computations have
not yet been tested experimentally.

\item Level-3 experiments consider situations in which the scale of a test
  is expanded into regimes in which 
various proposed alternative theories
predict an outcome other than that predicted by conventional physics.
The intent
   of such an experiment is to seek evidence for or against such an
   alternative theory.

 \item Level-4 experiments test physics in regimes for which there is not yet
   a
   standard theory which can be used to predict the outcome.  Instead we
   propose tests based on models which find their motivation from the
   expected (or guessed) nature of physics in these regimes.
Tests of quantum gravity fall into this latter
category.
\end{itemize}

\subsection{Summary}
\label{sec.summary}
We consider the experiments summarized in Table~\ref{tab.summary}.
For each experiment we indicate a characteristic length scale of the test,
the timeframe in which such an experiment may become technically feasible,
the domain of physics which is explored, a level classification as described in
Section~\ref{sec.classification}, and some indication of the magnitude of a
predicted effect.
In each case we put the smallest scale experiment considered --- many experiments can be
performed at larger scales as well.

\begin{table}
\centering
\begin{tabular}{|p{4.1cm}|p{1.7cm}|l|p{2cm}|c|p{3.2cm}|c|}
\hline
\textbf{Name} & \textbf{Scale} & \textbf{Timeframe} & \textbf{Regime} & \textbf{Level} & \textbf{Observability} & \textbf{Section}\\
\hline\hline
\multicolumn{4}{l}{\textbf{Entanglement Tests}}\\
\hline\hline
Long distance Bell-test & LEO and beyond & near-term & Standard QM & 1 & Observable & \ref{longdist}\\

Bell-test with human observers & 
Earth-Moon  & long-term & QM and free-will & 3 & Need human on Moon surface or lunar orbit & \ref{human.sec}\\

Detectors in relative motion & LEO & mid-term & Standard SR & 3 & achievable & \ref{moving_observers.sec}\\

Amplified entanglement & LEO & near--mid-term & QM & 3 & potentially achievable & \ref{macro_amp.sec}\\

Bimetric gravity & LEO & near-term & Test non-standard theory & 3 & bound parameter? & \ref{bimetric.sec}\\

\hline
\multicolumn{4}{l}{\textbf{Special and General Relativistic Effects}}\\
\hline\hline
Lorentz transformed polarization & LEO and beyond & mid-term & QM + SR & 1 & Beyond current tech. & \ref{lorentz-polar}\\

Relativistic frame dragging & TBD & TBD & QM + GR & 2 & Beyond current tech. & \ref{frame_dragging.sec}\\

Entanglement with curvature & TBD & visionary & QM + GR & 2 & Beyond current tech. & \ref{sec.parallel_transport}\\

Fermi problem & Sunshielded satellites & long-term & QFT & 2 & Borderline & \ref{Emm1}\\

Optical Co\-le\-lla-Over\-hau\-ser-Wer\-ner experiment & LEO and beyond & near-term & QM + GR & 1+ & Observable & \ref{cow.sec}\\

\hline
\multicolumn{4}{l}{\textbf{Accelerating Detectors in Quantum Field Theory}} \\
\hline\hline
Acceleration induced fidelity loss & TBD & visionary & QFT + GR & 2 & Beyond current tech. & \ref{unruh.sec}\\

Berry phase interferometry & LEO & mid-term & QFT + GR & 2 & Borderline & \ref{unruh.sec}\\

Gravitationally induced entanglement decorrelation & LEO and beyond
& near-term & Non-standard QFT + GR & 3 & Possibly observable & \ref{sec.Ralph}\\

Spacelike entanglement extraction & TBD & visionary & QFT + GR & 2+ & Beyond current tech. & \ref{EMM2} \\

\hline
\hline
\multicolumn{4}{l}{\textbf{Quantum Gravity Experiments}}\\
\hline\hline
Diffusion of polarization & TBD,  solar system? & visionary? & QG & 4 & Bound model parameters & \ref{diffusion.sec} \\

Spacetime noncommutativity & TBD & unknown & QG & 4 & unknown & \ref{noncommutativity.sec} \\

Relativity of locality & TBD,  solar system? & unknown & QG & 4 & unknown & \ref{DSR.sec} \\

\hline
\multicolumn{4}{l}{\textbf{Quantum Communication and Cryptographic Schemes}}\\
\hline\hline
Quantum tagging & LEO--GEO & near-term & QM + SR & 1 & feasible with current technology? & \ref{tagging.sec}\\

Quantum teleportation & LEO & near-term & Standard QM & 1 & feasible with current technology? & \ref{teleportation.sec} \\
\hline

\end{tabular}
\caption{Summary of possible experiments.
LEO refers to Low Earth Orbit, an elliptical
orbit about the Earth with altitude up to 2000 km.
The timeframes are mentioned in Section~\ref{timeframes.sec}.  Roughly, `near-term'
experiments ($\sim 5$ years) can
be accomplished with a single satellite in LEO, `mid-term' experiments
(25 years) require multiple satellites or higher orbits, `long-term' experiments involve
Earth-Moon distances, and `visionary' experiments extend to solar orbits and beyond.
Under ``Regime'' (and
throughout the paper) QM
refers to ordinary quantum mechanics, QFT to quantum field theory, SR to
special relativity, GR to
general relativity, and QG to quantum gravity.
The ``Level'' classifications are explained in Section~\ref{sec.classification}.
}
\label{tab.summary}
\end{table}

\section{Entanglement tests}
\label{sec.entanglement}

The space environment opens the possibility of performing entanglement
experiments over extremely long distances, allowing us to push the verification bounds
of quantum theory. The possibility of, thereby,
observing deviations from predictions of
the theory is tantalizing.

\subsection{Tests of local realism --- The ``Bell test''}\label{sec:belltest}

Quantum theory tells us that, within a multipartite entangled system, the
measurement-induced collapse of the state caused by measuring
one particle 
will be `instantaneously' reflected in measurement outcomes 
on the other particles, regardless of how far apart those particles may be. The troubling nature of this, first realized by Einstein, Podolsky, and Rosen~\cite{Einstein1935} (EPR), was later cast into the form of bipartite spin measurements by Bohm~\cite{bohm} (EPRB),
characterized rigorously 
by Bell~\cite{Bell64a}, and later again cast into an experimentally-testable manner by Clauser, Horne, Shimony, and Holt (CHSH)~\cite{Clauser1969}. Bell and CHSH each produced an inequality relating the statistical correlations of the outcomes of measurements performed on the two particles under two naturally intuitive assumptions: (1)~that physical systems possess only objective locally-defined properties (independent of measurement context), and (2)~that influences between systems cannot propagate faster than light-speed. Under certain configurations, quantum mechanics violates these ``Bell inequalities''.

Experimental tests of Bell inequalities involve gathering correlation statistics by measuring numerous entangled photon pairs. Thus far, these Bell tests have been consistent with quantum theory, and therefore, despite their intuitiveness, the assumptions of Bell's derivation do not hold.

\subsection{Long distance Bell test}\label{longdist}

The first experiment we consider involves
testing the phenomenon of quantum entanglement
over large distances. Quantum mechanics does not predict any breakdown in
its description of largely extended quantum states.  How valid is this assumption?
Current experiments on the ground have reached 144\,km, and distances beyond
that are difficult to impossible to be explored on the ground.
(There is also
 the possibility that quantum entanglement might be relevant
at cosmological scales~\cite{MartinMartinezMenicucci}, however it is not clear if observations or
experiments are possible to measure such entanglement.  See
Section~\ref{EMM2} for some discussion on this issue.)

Such an experiment would also have interest from the quantum foundations point of view: performing a long-distance Bell test with spacelike separated observers will lead to a paradoxical situation for some interpretations of QM when one considers the problem of measurement. For two spacelike separated events the concept of simultaneity is frame dependent and external observers in relative motion with respect to the experiment would answer differently to the question ``who measured first?''

As discussed in the work of Aharonov and Albert in the context of quantum
mechanics~\cite{AA81}, and later by Sorkin in the context of quantum field theory~\cite{Sorkin1993_proceeding}, assuming that quantum state reduction
upon measurement
takes place on hyperplanes
 can lead to causality paradoxes.  However, to the
authors' knowledge, an experiment testing for such a violation of causality 
has not
yet been performed.   Also, these results leave open other plausible
hypotheses about 
quantum state reduction --- in particular, that
its effects propagate causally~\cite{Kent-collapse,Kent-collapse09}.
We describe below (in Section~\ref{macro_amp.sec}) experiments testing this possibility.
The experimental challenges 
are
achieving relativistic relative velocities with respect to the EPR
experiment. The more spacelike separated Alice and Bob are, the easier the
simultaneity test is to perform. Satellite-based experiments can be key
in providing the necessary spacelike separation.

\begin{figure}[htbp]
   \centering
   \includegraphics[width=10cm]{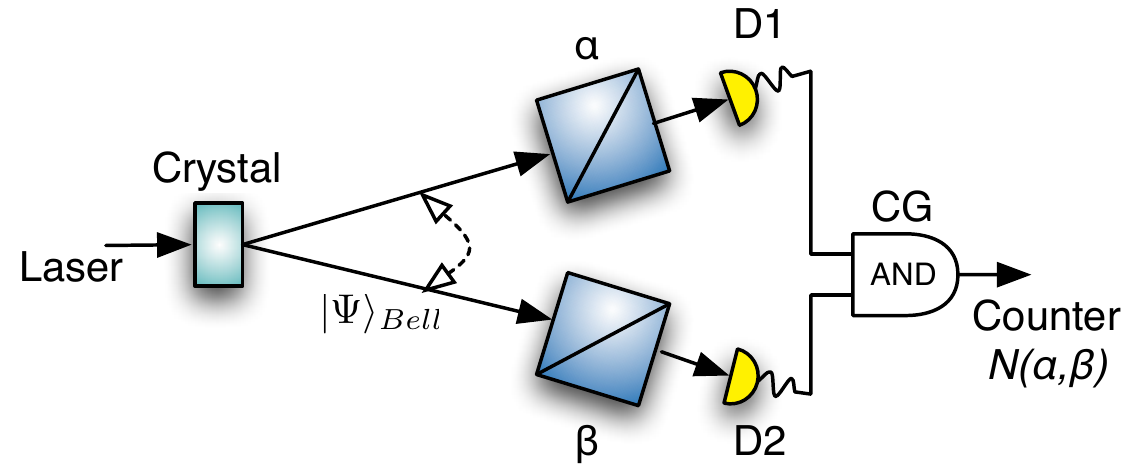} 
   \caption{Schematic view of a simple Bell-test experiment with entangled photons. Entangled pairs of photons are created in parametric down-conversion of a laser passing through an optical crystal. The entanglement properties of the detected photons are measured in two analyzers. Time-correlation of the photon detection signals is used to identify detections arising from photons that were generated as pairs, typically via a logical AND gate.
$\alpha$ and $\beta$ represent two possible measurement settings for each
detector, and $N(\alpha,\beta)$ the counts corresponding to each pair of settings.}
   \label{fig:belltest}
\end{figure}

A simple setup to test quantum entanglement for photon pairs is shown in
Figure~\ref{fig:belltest}.  A laser is sent through an optical crystal, which
creates pairs of photons possessing an entangled linear polarization state.
Each photon is sent through a polarizer, oriented at angles $\alpha$ and $\beta$, and then detected at D1 and D2 respectively. By tuning the orientations of the polarizers, it is possible to test the quantum correlations and observe if quantum entanglement violates local realism. (This correlation can also be utilized for quantum cryptography.)

In the case of ultra-long-range Bell tests, two alternative scenarios may be considered. The first is a symmetric setup, in which the quantum source is placed equidistant from two receivers. This is conceptually the most simple approach. A logistically easier (and therefore cheaper) alternative is to perform the experiment with one receiver, Bob, located at some distance, and a second receiver, Alice, at the same site as the entangled photon source. For example, this might represent satellite and ground receivers, respectively. Within this second scenario, however, if Alice's measurement is made on her photon immediately as it emerges from the source, then the two measurement events will not be spacelike separated, and the results from the experiment will be subject to locality and freedom-of-choice loopholes~\cite{scheidl_violation}.

To close these loopholes, one can insert a delay before Alice's measurement, such as a length of fibre-optic cable or a quantum memory device.  The appropriate choice of such a delay makes the measurement events spacelike separated, giving an experimental setup in which both the locality and freedom-of-choice loopholes can be closed while requiring only one measurement to take place at a distant site~\cite{scheidl_violation}.

\begin{figure}[htpb]
\includegraphics[width=\textwidth]{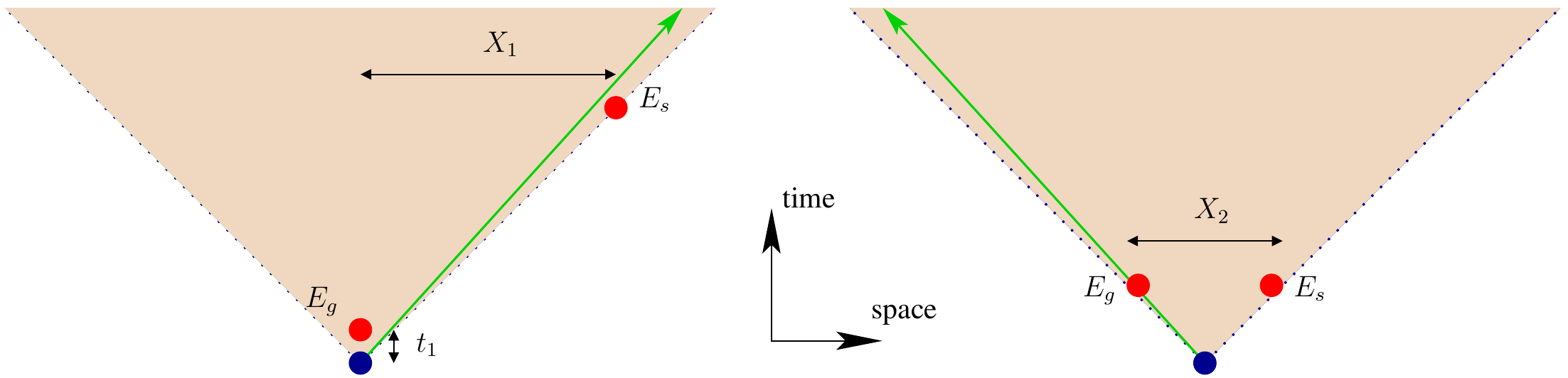}
\caption[Asymmetric Bell test]{Effective distance between receivers in an asymmetric Bell experiment.  Left: spacetime diagram of the experiment in Alice's frame, with two measurement events $E_g$ and $E_s$ on the ground (Alice) and on a satellite (Bob) respectively.  Right: spacetime diagram from perspective of reference frame in which the measurement events $E_g$ and $E_s$ are simultaneous. The effective spatial separation of the measurement events $X_2$ is much smaller than the altitude of the orbiting satellite. In each case the green arrow indicates the velocity of the other frame.}\label{fig.AsymmetricBell}
\end{figure}

Figure~\ref{fig.AsymmetricBell} illustrates such an ``asymmetric Bell test'' experiment, with a spacetime diagram drawn from the perspective of two reference frames.  The figure on the left is drawn in Alice's frame, the frame of a ground station in which the entangled photon source is located.  The emission event is colored in blue, and the two reception events are indicated by red dots, $E_g$ occurring at the ground station (Alice) a short time $t_1$ after the emission event, and $E_s$ occurring later on an orbiting satellite (Bob) at an altitude of $X_1$ above the ground station.

The figure on the right shows the same experimental scenario, from the perspective of a reference frame which is moving at a velocity $v$, away from the Earth's surface, such that the two detection events are simultaneous.  According to the right reference frame, the detection events occur at a spatial separation of $X_2$. The Lorentz invariant distance between the two events is
\begin{equation}
 \sqrt{(\Delta x)^2 - (c \Delta t)^2} = \sqrt{{X_1}^2 - [c (X_1/c - t_1)]^2} = X_2 \;.
\end{equation}
For a satellite orbiting at $X_1 = 1000$\,km and a quantum memory device which provides a $t_1 = 20\,\mu$s delay, this gives an effective separation of $X_2 = 109$\,km.  The two reference frames would be travelling at a relative speed of $v = \frac{X_1 - c t_1}{X_1} = 0.994 c$, which corresponds to a boost factor $\gamma = 83.6$.
For comparison, the Bell test detailed in Ref.~\cite{scheidl_violation},
for which the Earth-frame distance was 144\,km (and which 
employed 6\,km of optical fibre to yield a delay of 20\,$\mu$s) had a
corresponding 
Lorentz invariant separation of the detectors of 41\,km.

\subsection{Bell test with human observers}
\label{human.sec}

Leggett raises the concern that the 
locality loophole in Bell tests may not yet be completely closed
experimentally~\cite{Leggett:2009lr}.  He asserts that
\begin{quote}
``A truly definitive blocking of this loophole would presumably
require
that the detection be directly conducted by two human observers with a
spatial separation  such that the signal transit time exceeds human
reaction
times, a few hundred milliseconds (i.e.\ a separation of several tens of
thousand kilometers).  Given the extraordinary progress made in quantum
communication in recent years, this goal may not be indefinitely far
in the future.''
\end{quote}

A Bell experiment employing human observers could be performed at the scale
of the Moon's orbit about the Earth.  For example, one observer may sit in a
lab on the lunar surface, while another sits on the Earth's surface, with an
entangled photon source located in a satellite at high Earth orbit.
The possibility of employing humans to decide on the measurement setting in a
Bell test leads to interesting tests of the role of free-will in quantum mechanics, as mentioned in Ref.~\cite{Kaiser2006Gravitat}.

\subsection{Bell test with detectors in relative motion}
\label{moving_observers.sec}

This type of experiment is of interest for its relevance to the foundations of quantum
mechanics. As already discussed in Section~\ref{longdist}, some
interpretations of quantum mechanics 
impart ontological physical meaning to wavefunction collapse.
For such interpretations, as
discussed above, the concept of quantum state reduction can lead to problems when relativity is considered, in particular with situations in which spacelike separated measurements take place~\cite{AA81,Sorkin1993_proceeding}.
Other interpretations, such as the ensemble interpretation~\cite{BallentineB} or consistent histories~\cite{DowkerKent} will be perfectly compatible with the expected outcomes of these experiments~\cite{P00b}.

\begin{figure}[hbt]
    \centering
    \includegraphics[width= 6cm,angle=-90]{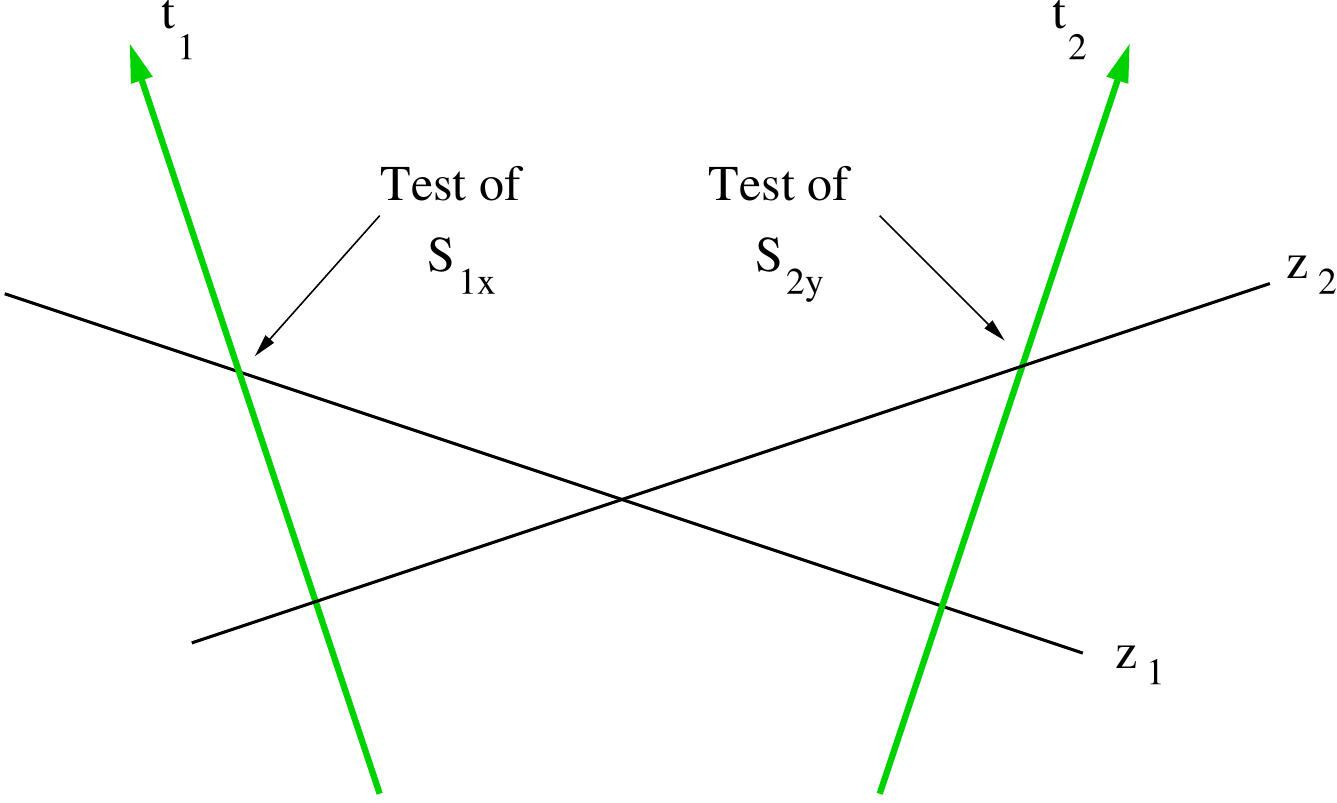}
    \caption[Peres Timing Paradox]{ From Ref.~\cite{peres-b}, this spacetime diagram shows the coordinate systems and the locations of the two tests, $S_1$ and $S_2$. The $t_1$ and $t_2$ axes are the world lines of observers who are receding from each other. In each Lorentz frame, the $z_1$ and $z_2$ axes are isochronous: $t_1=0$ and $t_2=0$, respectively.} \label{fig.PeresTimingParadox}
\end{figure}

Space platforms allow for (and in some situations, mandate) relative motion between a pair of observers at different locations with considerably less restriction than on Earth. By exploiting this, one is able to test interpretations of wave function collapse in scenarios for which the two observers disagree on the relative time ordering of the measurement events.
If the two distant observers measure the photons, events $S_1$ and $S_2$, under a large relative velocity and spatial separation, then due to special relativity their lines of instantaneity (isochronous planes) will become shifted by a measurable amount. 
This is a relativistic generalization of Bohm's version of the Einstein-Podolsky-Rosen ``paradox'';
see Ref.~\cite{peres-b} (Section~6.1) and Figure~\ref{fig.PeresTimingParadox}.
Due to the mutual time shifts, the two observers may each measure their particle later than the other (an ``after-after'' scenario) in the case that the observers move apart from each other, or each measure their particle earlier than the other (``earlier-earlier''), in the case of approaching motion.
Such experiments were first proposed in 1997 by Suarez and Scarani~\cite{scarani97}.

One picture of entanglement has it that the ``first'' measurement influences the outcome of the ``second'' measurement, perhaps non-locally. However, this 
picture breaks down in a situation in which 
special relativity mandates that the time-ordering of events is ambiguous.
The probabilities predicted by quantum theory do not depend on the
time-ordering of spacelike events, so its predictions will not be changed. Yet this \emph{timing paradox} leaves us with some puzzles about understanding the physical reality  of quantum states and the non-local collapse of the wave functions. Essentially, in  such an experiment we cannot even objectively define the non-local update of one system because of the measurements performed on the other.
It is important to note that it is the measurements ---  external interventions~\cite{P00b} into the system --- and not entanglement that are responsible for the ``paradox''. If a pair of spin-half particles is prepared in a direct product state and the local measurements on the particles are  mutually spacelike, then  the state of the system does not have a Lorentz-covariant transformation law outside the common past and the common future of these interventions~\cite{pt02}.
The Liouville function of two classical particles that are subject to two mutually spacelike stochastic events has the same property~\cite{pt02}.

With space platforms it is possible to reach relative velocities and spatial separations which allow this ``paradoxical'' situation to be observed and explored.
The respective time shift between the two observers by the Lorentz transformation is
\begin{equation}
t' = \frac{1}{\sqrt{1 - v_0^2 / c^2}} \left(t - \frac{v_0 x}{c^2}\right),
\end{equation}
and by setting the origin of reference frame $t'$ to be zero, we obtain:
\begin{equation}
t = \frac{v_0 x}{c^2}.
\end{equation}
As an example, by inserting the velocity of $2 \times 7.5$\,km/s = 15\,km/s for the relative motion of two LEO satellites, we obtain the following temporal shift per kilometer:
\begin{equation}
t/x =  \frac{v_0 }{c^2} =  \frac{15\,\mathrm{km/s}}{(3 \times 10^{5}\,\mathrm{km/s})^2} =   166\,\mbox{ps/km}
\end{equation}
Assuming fast random number generators and optical switches allow randomly setting the analyzers every 10\,ns (which is more than 10 times faster than reported in Ref.~\cite{scheidl_violation}), this would require a minimal separation of the two satellite observers of about 60\,km. Assuming the measurements can be performed up to the maximal separation for  two satellites in LEO of about 1500\,km, the time available for measurements would be about one and a half minutes,
which should be sufficient for a valid Bell test.

Such an experiment comes with another important requirement, namely that the emission of photon pairs from the source must be very well timed, so that the two photons travel exactly the same time from the source to their respective measurement stations. The synchronicity requirement is better than 10\,ns for the shortest separation of 60\,km, and expands up to 250\,ns at 1500\,km, which is challenging but seems achievable.

\subsection{Bell experiments with macroscopic amplification}\label{macro_amp.sec}

At first sight, the quantum projection postulate tells us that something
alters when a measurement takes place: the state vector of the
measured quantum system changes discontinuously as a result of the
measurement.   Of course, one view of quantum theory, in line with
several different interpretational ideas, is that
this ``collapse of the wave function'' is merely a mathematical
operation that physicists carry out, and does not represent
any real physical process.  While that view may turn out to be
correct, it presently leaves us with some fundamental
puzzles.   If we cannot extract a description of physical reality
from the quantum wave function, then how \emph{do} we find a
description of physical reality within quantum theory?  Or if we can't,
how do we make sense of a fundamental physical theory that
doesn't describe an objective reality (and so doesn't seem to
describe observers or measuring devices either)?    But if
we \emph{can} extract a description of physical reality from
the wave function, why wouldn't we expect that reality to change
discontinuously when the wave function does?

On the other hand, the idea that wave function collapse might be
a real physical process also runs into immediate difficulties, one
of which is that, as normally formulated, the description of
wave function collapse depends on one's choice of reference frame,
and so taking it to be real seems to conflict with special
relativity.

Reviewing all the arguments and counter-arguments and the
different interpretational ideas supporting each is
beyond our scope here.  We simply want to
note an interesting non-standard line of thought about wave function
collapse that motivates new experimental tests of quantum theory.

To motivate this, we first suppose that
wave function collapse \emph{is} a real objectively defined process,
and moreover
one that is \emph{localized}: collapses occur as the result of
events that take place at definite spacetime points.
On this view, collapses are not fundamentally defined in
terms of observers or measurements.
The collapse process obeys some as yet unknown
mathematical laws --- perhaps, for example,
some relativistic generalization
of the Ghirardi-Rimini-Weber-Pearle (GRWP)~\cite{Ghirardi86a,GPR} spontaneous collapse
models --- from which the usual account of collapses taking place as
the result of measurements in our experiments emerges as a special
case and an approximation.

We also suppose that collapses propagate in a way that respects
special relativity.   A collapse event at a spacetime point $P$ only
affects physics in the future light-cone of $P$: in particular,
it does not affect the probabilities of any other collapse events
at points spacelike separated from $P$.

Taken together, these hypotheses give a way of defining an alternative
to standard quantum theory, which we call \emph{causal quantum theory},
that makes very different predictions from standard quantum
theory~\cite{Kent-collapse,Kent-collapse09}.   To define causal quantum
theory precisely,
we would need a precise formulation of the
dynamical collapse laws, which needs a relativistic collapse theory.
(So causal quantum theory is really an umbrella
term for a class of theories.)
But we can get some intuition about the nature of the alternative by
supposing that the existing (non-relativistic) GRWP models are a good
approximation to this unknown relativistic theory for experiments
where relativistic quantum effects are relatively negligible.
In particular, we can look at the implications for Bell experiments.

Causal quantum theory predicts that, in a Bell test in which
measurements on two separated particles in a singlet state
are genuinely completed (i.e.\ the relevant underlying collapse
events take place) in spacelike separated regions, essentially \emph{no}
correlations will be observed in the measurement outcomes.
Observing spin up on one wing does not affect the probabilities of
observing spin up or down on the other.   This is, of course, very
different from the correlations predicted by standard quantum theory
and apparently observed in essentially all Bell experiments to date.
It leads to predictions of outcomes that are very unlikely, according
to standard quantum theory.
These predictions nonetheless never actually cause a logical
contradiction: causal quantum theory is self-consistent~\cite{Kent-collapse,Kent-collapse09}.

At first sight, causal quantum theory may seem to be clearly refuted by the
overwhelming experimental evidence for Bell correlations.
However, this in fact raises a new worry about all existing
Bell tests: (how) \emph{do} we know that collapse events sufficient
to effectively define an irreversible measurement do in fact take
place in spacelike separated regions in the two wings of the
experiment?   Is there a possible alternative, namely
that the wave function remains uncollapsed
until a later point in the measurement chain, when the signals
produced in the two wings are brought together and compared?
If so, causal quantum theory and standard quantum theory would
then both predict the standard quantum correlations actually
observed.  But then no Bell experiment to date would actually have
succeeded in demonstrating quantum non-locality.
After all, causal quantum
theory is, as its name suggests, a locally causal theory.
If it is consistent with the results of any given experiment, then by definition
this experiment cannot have demonstrated non-locality in nature.

Quantitative estimates suggest this is indeed a live issue,
at least in principle.
With an appropriate choice of collapse parameters, GRWP models
would indeed predict that there are essentially no collapses
in spacelike separated regions on the two wings of standard
Bell test experiments.   For example, neither the avalanches of
photons generated by detecting a photon in a photo-multiplier,
nor the electrical currents that ensue, are sufficiently
macroscopic to (necessarily) correspond to a GRWP collapse
in anything like a sufficiently short time.   There is a loophole
--- the \emph{collapse locality loophole} --- in all Bell experiments
to date.  One can find collapse models, and so versions of causal
quantum theory, in which none of these experiments would (essentially)
ever cause spacelike separated collapse events in the two wings.
No collapse occurs, and so no definite outcomes or correlations are
generated, until after the signals are brought together and compared.
If this description were correct, quantum non-locality would never actually
have been demonstrated.

To close this loophole, one needs to carry out Bell tests
in which the measurements in the two wings produce
\emph{macroscopically} distinguishable outcomes in spacelike separated
regions.  In principle, one simple way to do this, in line with the GRWP model
and also
with Penrose~\cite{penrose} and Diosi's~\cite{diosi} intuitions
regarding
a link between gravitation and collapse, is to arrange experiments so that massive objects are
quickly moved to different positions depending on the measurement
outcomes.    The technological challenge
of arranging for a mass to move as the result of a measurement over
timescales short compared to terrestrially achievable Bell experiment
separations is, however, considerable.

Nonetheless, Salart et~al.~\cite{SBHGZ08} were able to
carry out a beautiful terrestrial experiment, motivated by these
ideas, exploiting the rapid deformation of piezocrystals when
a signal voltage is applied.
In their experiment, a gold-surfaced mirror measuring $3 \times 2
\times 0.15$\,mm and weighing $2$\,mg is displaced by a distance
of at least $12.6$\,nm by a piezocrystal activated by a detector
in a Bell experiment.  The displacement is complete within
$6.1\,\mu$s after the photon enters the detector.
Using estimates due to Penrose~\cite{penrose} and Diosi~\cite{diosi}
for a plausible collapse time if the collapse is related to the
gravitational energy of the mass distribution associated with the
two superposed mirror states, Salart et~al.\ obtained a collapse time
of 1\,$\mu$s, suggesting that the measurement is complete and
a collapse has taken place within 7.1\,$\mu$s.  The two
wings of their experiment were separated by 18\,km, i.e.\ $c \times
60\,\mu$s, so that on this interpretation the collapses in the
two wings are spacelike separated.   As their results agreed
with standard quantum theory, they refute causal quantum theory for a
range of collapse model parameters.

That said, neither Penrose nor Diosi's estimates derive from a
consistent theory of gravity-related collapse.  There are
a variety of ways to try to build such models (see, e.g., Refs.~\cite{ggr} and~\cite{pearlesquires}).
The predicted collapse rates in any given
experiment are very sensitive to details of such models,
which are not currently fixed by any compelling theoretical principle.
Moreover, the earlier GRWP collapse models~\cite{Ghirardi86a,GPR} (which do not link collapses
directly with gravity) and other types of collapse model
are also well-motivated.   For these models too, the
predicted collapse rates in any given experiment are very
sensitive to uncertain details and ad-hoc choices.
For example, one possible extrapolation~\cite{kentgrw} of the original
GRW collapse model to indistinguishable particles would suggest
a collapse time of ${>}10^2$\,s for the mirror superposition states
in the Salart et~al.\ experiment.

The Salart et~al.\ experiment has thus by no means closed the
collapse locality loophole completely.  To do that, we would need
to arrange a Bell experiment in which different measurement
outcomes produce matter distributions so macroscopically
distinct, and in which the separation between the wings
is long enough in natural units, that no remotely plausible
collapse model can predict that the systems on the two
wings remain in superposition throughout the experiment.

This can and presumably ultimately will be achieved by Bell
experiments based in deep space, with very large separations
between the wings.
Interestingly, though, even near-Earth experiments can achieve
a great deal.  A Bell experiment on a scale of $10^3$--$10^5$\,km
would be able to exclude any model that predicts collapse times
shorter than ${\approx}$ 3--300\,ms.  The upper end of this range compares
well with human reaction times (${\approx} 100$\,ms); even the lower
end compares well to the time required to create an effect on
a scale of meters at explosive detonation velocities
(${\approx} 10^3$--$10^4$\,m/s).
To have any  plausible motivation, a collapse model must produce collapses of
macroscopic superpositions perceptible by humans on timescales
short compared to those we can discriminate.   A model which fails to do
this has to explain why we see definite outcomes while being, at least
for some perceptible length of time, actually in indefinite
superpositions.

Some creative engineering ingenuity is
required to devise macroscopic amplifications well suited for near-Earth
Bell experiments, given the constraints of
practicality and repeatability.   Mechanical amplifications of a
detector signal to move macroscopic masses are presumably more
practical and more efficient than those relying on human intervention,
and more repeatable than those involving explosives.
For the moment, we offer this design problem as a challenge to
experimental colleagues, encouraged by the above figures,
which suggest that even near-Earth, well-designed experiments should be able
to close the collapse locality loophole very significantly, and perhaps
indeed completely.

\subsection{Bimetric gravity}\label{bimetric.sec}

Moffat has questioned whether quantum entanglement is
divorced from our common intuitive ideas about spacetime and
causality, and 
proposed a
relativistically causal description of quantum entanglement in terms of a
bimetric theory of gravity~\cite{moffat}.
Such a theory should be testable with a Bell experiment in which the two
receivers are moving relativistically with respect to each other, as
described in Section~\ref{moving_observers.sec}.

Quantum entanglement, according to the standard interpretation, is a \emph{purely quantum phenomenon} and classical concepts associated with causally connected events in spacetime are absent. This is the point of view promulgated by
standard quantum mechanics.
One should abandon any notion that physical space plays a significant
role for distant correlations of degrees of freedom associated with particles
in entangled quantum states.  For those who remain troubled by this
abandonment of a spatial connection between entangled states, it is not clear
how attempting to change quantum mechanics would help matters. This leaves
the possibility that classical special relativity is too restrictive to allow
for a complete spacetime description of quantum entanglement. An alternative
scenario is based on a ``bimetric'' description of spacetime. In a bimetric
theory, the light cones of two metrics describe classical and ``quantum''
spacetimes~\cite{moffat}. The quantum spacetime is triggered by a measurement
of a quantum system and allows the propagation of quantum ``information'' at
superluminal speeds prohibited by the lightcone in the classical spacetime.

The satellite experiment proposed in Section~\ref{longdist} may already be
sufficient to provide evidence for 
bimetric scenarios.
For example, in the case that the ``quantum mechanical metric'' of Ref.~\cite{moffat} has a Lorentzian signature (as is depicted in Figure~1 of Ref.~\cite{moffat}) then the two measurement events at
$E_g$ and $E_s$ of Figure~\ref{fig.AsymmetricBell} must be causally related
by the quantum metric if quantum correlations are observed.  It is possible that by changing the delay on the
ground, or by a changing distance to a satellite in an eccentric orbit, the
measurement events may cease to be in causal contact, according to the
quantum metric, in which case the quantum correlations predicted by the
theory would vanish.

\section{Relativistic effects in quantum information theory}
\label{sec.entanglementSRGR}

With the possible exception of Section~\ref{sec:q_grav_expts},  spacetime in our
discussion is either (weakly) curved or an (approximately) flat fixed background for propagation of polarization qubits.
The simplest approximation that is tacitly assumed in most of  the discussions of long-distance quantum information processing is to treat photons as massless point particles that move on the rays prescribed by geometric optics (e.g.,
 null geodesics in vacuo) and carry transversal polarizations. The latter can
 be described either in terms of a two-dimensional Hilbert space or a complex
 three-dimensional vector, which is orthogonal to the photon's
 momentum~\cite{PT03, T10lnp} (Section~\ref{lorentz-polar}). Evolution of a
 polarization  along the ray is  described by the first post-eikonal
 approximation~\cite{BWopt}, and the resulting rotation is interpreted as
 quantum phase. Validity of this approximation is justified because of the
 scales involved.  The finite-width wave packet effects, while introducing many interesting features, are of higher order~\cite{PT03, LPT}.

Therefore, photons follow null trajectories in spacetime with    tangent four-vectors $k$, $k^2=0$, and the spacelike polarization vectors $f$, $f^2=1$, are
 parallel-transported along the rays~\cite{mtw,chandra},
\beq
{\anab_k} k=0, \qquad {\bnab}_k f=0, \label{ray}
\eeq
where ${\anab}_k$ is a covariant derivative along $k$.

All experiments which contain at least one receiver in orbit
will involve 
high-precision reference frame alignment, in a general relativistic setting.
This raises some interesting issues, as discussed below.

\subsection{Special relativistic effects}
\subsubsection{Lorentz transformations and polarization}\label{lorentz-polar}

Every  Lorentz transformation $\Lambda$ that connects two reference frames  results in a unitary operator that
 connects the two  descriptions of a quantum state,
$|\Psi'\9=U(\Lambda)|\Psi\9$~\cite{wei1}. The unitaries $U(\Lambda)$ are obtained using
Wigner's induced representation of the Poincar\'{e} group~\cite{wkt,wei1}. The first step in its construction is also used to build a polarization basis in a stationary curved spacetime (Section~\ref{frame_dragging.sec}).

Single-particle states
belong to some irreducible representation. Two invariants --- the mass $m$ and the intrinsic spin
$j$ --- label
the representation. The basis states are labelled by three components of the
4-momentum $p=(p^0,\bp)$ and the spin along a particular direction. Hence a generic
state is given by
\begin{equation}
|\Psi\9=\sum_\sigma\!\int \!d\mu(p)\psi_\sigma(p)|p,\sigma\9,
\end{equation}
where $d\mu(p)$ is the Lorentz-invariant measure and the
momentum and spin eigenstates are $\delta$-normalized
and are complete on the one-particle Hilbert space.

The single-photon states are labelled by momentum $\bp$  and helicity
$\sigma_\bp=\pm 1$, so the state with a definite momentum is given
by $\sum_{\sigma=\pm 1}\alpha_\sigma|p,\sigma_\bp\9,$ where
$|\alpha_+|^2+|\alpha_-|^2=1$. Polarization states are also labelled
by 3-vectors $\bep^\sigma_\bp$, $\bp\cdot\bep_\bp^\sigma=0$, that
correspond to the two senses of polarization of  classical
electromagnetic waves. An alternative labelling of the same state,
therefore, is $\sum_{\sigma=\pm
1}\alpha_\sigma|p,\bep^\sigma_\bp\9$~\cite{PT04,T10lnp}.

\begin{figure}[!phtb]
\begin{center}
 \vspace{-7. cm}
\includegraphics[width=0.65\textwidth]{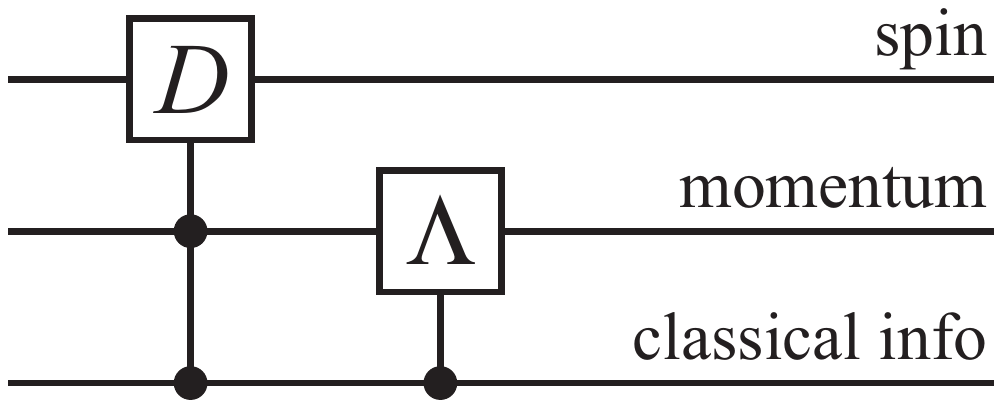}
\vspace{-7.5cm}
\caption{Relativistic state transformation as a quantum circuit: the gate $D$
which represents the matrix $D_{\xi\sigma}[W(\Lambda,p)]$ is
controlled by both the classical information and the momentum $p$,
which is itself subject to the classical information (i.e.\ the Lorentz transformation $\Lambda$ that relates the reference frames).}
\label{fig-lor}
\end{center}
\end{figure}

Action of  the unitary operator
$U(\Lambda)$ is represented  in Figure~\ref{fig-lor}.
It is derived with respect to the standard
4-momentum, which for photons is taken to be $k_R=(1,0,0,1)$, and the standard Lorentz transformation
\beq
L(k)=R(\hbk)B_z(u),
\eeq
 where $B_z(u)$ is a pure boost along the $z$-axis with a
velocity $u$ that takes $k_R$ to $(|\bk|,0,0,|\bk|)$ and $R(\hbk)$
is the standard rotation that carries the $z$-axis into the
direction of the unit vector $\hbk$~\cite{wei1}. If $\hbk$ has polar and
azimuthal angles $\theta$ and $\phi$, the standard rotation
$R(\hbk)$ is accomplished  by a rotation by $\theta$ around the
$y$-axis, followed by a rotation by $\phi$ around the
$z$-axis.

Explicitly, the transformation is given by
\beq
U(\Lambda)|p,\sigma\9=\sum_\xi D_{\xi\sigma}[W(\Lambda,p)]|\Lambda
p,\xi\9,
\eeq
where $D_{\xi\sigma}$ are the matrix elements of the representation
of  the Wigner little group element $W(\Lambda,p)\equiv L^{-1}(\Lambda p) \Lambda L(p)$ that
leaves $k_R$ invariant, $k_R=Wk_R$.

Helicity is invariant under proper Lorentz transformation, but the
basis states acquire phases.
An arbitrary little group element for a massless particle is
decomposed according to $W(\Lambda,p)=S(\beta,\gamma)R_z(\xi),$
where the elements $S(\beta,\gamma)$ do not correspond to the
physical degrees of freedom~\cite{wei1}. As a result, the little group elements are represented
by
\beq
D_{\sigma'\sigma}=\exp(i\xi\sigma)\delta_{\sigma'\sigma}, \qquad \sigma=\pm 1. \label{dss}
\eeq

For pure rotations,
$\Lambda=\eR$ (in this case the same letter is used for both a 4D matrix and its non-trivial 3D block), the phase has a particularly simple form~\cite{PT03,LPT}.  We
  decompose an arbitrary
rotation $\eR$, $\bk=\eR\bp$, as
\beq
\eR=R_{\hbk}(\chi)R(\hbk)R^{-1}(\hbp), \label{omdef}
\eeq
where $R_{\hbk}(\chi)$ characterizes a rotation around $\hbk$, and
$R(\hbk)$ and $R(\hbp)$ are the standard rotations that carry the
$z$-axis to $\hbk$ and $\hbp$, respectively. Then
\beq
W(\eR,p)=R_z(\chi).\label{wr}
\eeq

A practical description of polarization states is given by spatial
vectors that correspond to the classical polarization directions.
Taking again $k_R$ as the reference momentum, two basis vectors of
linear polarization are $\bep_{k_R}^1=(1,0,0)$ and
$\bep_{k_R}^2=(0,1,0)$, while to the right and left circular
polarizations correspond $\bep_{k_R}^\pm=(\bep_{k_R}^1\pm
i\bep_{k_R}^2)/\sqrt{2}$.

Phases of the states  obtained by the standard Lorentz
transformations $L(k)$ are set to 1. Since the standard boost
$B_z(u)$ leaves the four-vector $(0,\bep^\pm_{k_R})$ invariant, we
define a polarization basis for any $\bk$ as
\beq
\bep^\pm_\bk\equiv R(\hbk)\bep^\pm_{k_R}\quad \Leftrightarrow\quad |k,\pm\9\equiv L(k)|k_R,\pm\9, \label{conve}
\eeq
while the transformation of polarization vectors under an arbitrary $\eR$ is given by the rotation itself. Indeed,
\beq
\eR\bal(\bp)=R_{\hbk}(\chi)\bal(\hbk) \quad \Leftrightarrow \quad U(\eR)(\alpha_+|p,+\9+\alpha_-|p,-\9)=
\alpha_+e^{i\chi}|k,+\9+\alpha_-e^{-i\chi}|k,-\9.
\eeq
For a general Lorentz transformations the triad
$(\bep^1_\bp,\bep^2_\bp, \hbp)$ is still rigidly rotated, but in a more
complicated fashion~\cite{T10lnp, LPT,caban}.

Since LEO satellite velocities are  of the order $v/c\sim 10^{-5}$ the first order expansion for a pure boost~\cite{caban} is sufficient to estimate the induced phase. If the photon propagates in the first frame along $\hbp=(\sin\theta\cos\phi,\sin\theta\sin\phi,\cos\theta)$ (we assume $\theta<\pi/2$) and the frames are related by the boost $v(\sin\theta_b\cos\phi_b,\sin\theta_b\sin\phi_b,\cos\theta_b)$, then
\beq
\chi=-\frac{1}{2}\tan{\half\theta} \sin\theta_b \sin(\phi-\phi_b)\frac{v}{c}.
\eeq

The effect is most pronounced ($\chi\simeq-\half v/c\sim 10^{-5}$) when the light is sent in a direction perpendicular to the reference $z$-axis, with the receiver velocity being perpendicular to both.

The Lorentz transformation also influences the diffraction angle for wave packets~\cite{PT03}. For a detector moving with respect to the emitter with velocity $v$ along the propagation direction of a narrow beam ($\theta \ll v/c$), the diffraction angle changes to
\beq
\theta'=\theta\sqrt{\frac{1+v/c}{1-v/c}}.
\eeq

\subsubsection{Entanglement}
\label{ent-sr}
As long as finite wave-packet width effects can be ignored~\cite{PT03} a bipartite or multipartite polarization entanglement is preserved. This can be seen as follows. Lorentz boosts do not
create spin-momentum entanglement when acting on
eigenstates of momentum, and the effect of a boost on a
pair is implemented on both particles by local unitary
transformations, which are known to preserve entanglement. This conclusion is valid for both massive and massless particles. Since the momenta are known precisely, the phases that polarization states acquire are known unambiguously, and in principle can be reversed. Hence, for a maximally entangled pair in the laboratory frame, the directions of perfect correlation for two photons
still exist in any   frame, even if they are
different from the laboratory directions~\cite{hacyan, tera03}. Effects of the finite wave-packet spread  on the observation of entanglement with  moving detectors were studied in Ref.~\cite{gba03,lmt10}.

\subsection{General relativistic effects}

\subsubsection{Relativistic frame dragging}
\label{frame_dragging.sec}

A somewhat imprecise  term ``frame dragging''~\cite{inertia, sted03,iorio,will} refers to a number of phenomena that, in a   field of a  slowly
rotating mass,  can be attributed to the presence of the mixed spacetime  components  of the metric $h_i=\sg_{0i}$~\cite{mtw,inertia}. In contrast to geodetic effects that result from mass-energy density, the former are  induced by the mass-energy currents of the source. The field $h_i$ plays a role analogous to the vector potential of electromagnetism. We refer to these and similar phenomena as gravitomagnetic~\cite{inertia}.  A  frame-independent indicator of gravitomagnetism is a non-zero value of the pseudo-invariant $^*R_{\kappa\lambda\mu\nu}R^{\kappa\lambda\mu\nu}$ (where $^*R_{\kappa\lambda\mu\nu}$ is a dual of the Riemann tensor), which for an isolated system is proportional to its angular momentum $J$~\cite{inertia}.

Two of the best-known gravitomagnetic effects --- Lense-Thirring precession  of the orbit of a test particle and Schiff precession of the axis of a gyroscope ---
 were   used in precision tests of general relativity~\cite{willn,lageos, gpb}. In the near-Earth environment the effects are small and challenging to measure: gravitomagnetic  precession rates for the LAGEOS experiment (Lense-Thirring effect) and the Gravity Probe-B (Schiff precession)   are 31\,arc\,msec/yr and 39\,arc\,msec/yr, respectively.

 Gravitomagnetism   also influences   propagation and polarization of  electromagnetic waves. Changes in polarization (a gravitomagnetic/Faraday/Rytov-Skrotski\u{\i} rotation~\cite{skrot57}) are operationally meaningful only with respect to a  local polarization basis~\cite{BT11}.

A reference-frame term~\cite{BT11, BDT11} contributes to the polarization rotation alongside the Machian gravitomagnetic effect~\cite{god70,fayos}.
In a stationary spacetime the  Landau-Lifshitz 1+3  formalism~\cite{ll2} lets us  rewrite the 4D parallel transport equations  in a 3D form~\cite{fayos}, demonstrating  the joint rotation of the unit polarization $\hbf$  and the unit wave vector $\hbk$~\cite{BDT11},
\beq
\frac{D\hbk}{d\lambda}=\Omegab\times\hbk, \qquad
\frac{D\hbf}{d\lambda}=\Omegab\times\hbf, \label{3dprot}
\eeq
where $D/d\lambda$ is a 3D covariant derivative (with respect to an affine parameter $\lambda$). The angular velocity of rotation $\Omegab$ is given by
\beq
\Omegab=2\omegab-(\omegab\!\cdot\!\hbk)\hbk-\bE_g\times\bk.
\eeq
Here the gravitoelectric term $\bE_g$
reduces to the Newtonian gravitational acceleration in a non-relativistic limit, and $\omegab=\mathbf{B}_g$/2 is a gravitomagnetic term.

Transversality of electromagnetic waves determines the plane of polarization for each propagation direction, but   a choice of a polarization basis is still free. Only by referring to these bases can the evolution of $\hbf$ be interpreted as polarization rotation. In Minkowski spacetime this choice is made once for all spacetime points, as in Section~\ref{lorentz-polar}. Comparisons between different points on a curved background require connections.

Consider some choice of a (linear) polarization basis $\hbb_{1,2}(\hbk)$ along a photon's path.
 By setting $\hbf=\hbb_1$ at the starting point of the trajectory, the phase $\chi$ appears as $\hbf(\lambda)=\cos\chi\hbb_1+\sin\chi\hbb_2$, so
\beq
\frac{d\chi}{d\lambda}=\frac{1}{\hbf\!\cdot\!\hbb_1}\left(\frac{D\hbf}{d\lambda}\!\cdot\! \hbb_2+\hbf\!\cdot\!\frac{D\hbb_2}{d\lambda}\right)=\omegab\!\cdot\!\hbk+\frac{1}{\hbf\!\cdot\!\hbb_1}\hbf
\!\cdot\!\frac{D\hbb_2}{d\lambda}. \label{defrot}
\eeq
The term $\omegab\!\cdot\!\hbk$ is the Machian contribution~\cite{god70}, but the reference-frame term may be equally or even more important~\cite{BT11}.

A commonly held view that no polarization phase is accrued in the Schwarzschild spacetime can be substantiated in the so-called Newton gauge~\cite{BT11}, where the fiducial $z$-axis (Section~\ref{lorentz-polar}) is chosen along the free-fall acceleration as seen by a static observer, and the standard polarization direction $\hbb_2$ is directed along $\hbz\times\hbk$.
In any spacetime, a phase accumulated along a closed path is gauge-independent  and can be expressed as an area integral of an appropriate curvature~\cite{BDT11}.

Two examples (in the Newton gauge) will illustrate the order of magnitude of the effects. In the Kerr spacetime, the calculations are simplified by having a sufficient number of conserved quantities~\cite{chandra,ll2}.
Outgoing geodesics in the principal null congruence satisfy a number of remarkable properties (including that these are trajectories of a constant spherical angle $\theta$) and can be easily integrated~\cite{chandra}. The leading order of polarization rotation is found by a direct integration,
\beq
\sin\Delta\chi=-\frac{J}{Mc}\left(\frac{1}{r_1}-\frac{1}{r_2}\right)\cos\theta,
\eeq
where $r_1$ and $r_2$ are initial and final radial coordinates, respectively. The Earth's angular momentum is $J_\oplus=5.86\times 10^{40}$ cm$^2$g/s  and its mass is $M_\oplus=5.98\times 10^{27}$\,g.
Sending a photon  on such a trajectory, starting at  $r_1=\mbox{12,270}$ km (the
semimajor axis of the LAGEOS satellite orbit~\cite{lageos}), at the latitude
of $45^\circ$ and taking $r_2\rightarrow\infty$, we obtain the rotation of $\Delta\chi\sim
39$\,arc\,msec in a single run, and nearly twice this amount if the light is sent from the ground. This result, however, depends on the special
initial conditions~\cite{BT11}.

The results typically scale as the inverse square of a typical minimal distance, $\Delta\chi\propto r_{\mathrm{typical}}^{-2}$. For example~\cite{BT11,BDT11} a photon that was emitted and detected  far from the spinning gravitating body, but passed close to it.  In a special case of  emission along the axis of rotation, we get in the leading order
\beq
\sin\chi=\frac{4GJ}{s^2c^3}, \label{chi1}
\eeq
where  $s$ is the impact parameter~\cite{BDT11}. The antiparallel initial direction gives the opposite sign. For the Earth (and the impact parameter being its radius), the resulting phase is minuscule: $3\times 10^{-7}$\,arc\,msec.

The above examples provide an estimate for the order of magnitude of a
potentially observable effect, however they are not gauge invariant.  The
latter requires the photons to traverse a closed path.
While such an experiment does not require a complicated alignment of the polarizers, at least three nodes are necessary to produce a path that encloses a non-zero area. The conceptually simplest  path is formed in the following setting. The trajectory  starts parallel to the axis of rotation with an impact parameter $s_1$ (and the initial angle $\theta_1=\pi$). Far from the gravitating body (so its influence can be ignored), the outgoing photon is twice reflected and sent in again with the impact parameter $s_2$ and the initial angle $\theta_2=0$. After the second scattering   and appropriate reflections it is returned to the initial position with the initial value of the momentum. Then
\beq
\Delta\chi=\frac{4GJ}{c^3}\left(\frac{1}{s_1^2}-\frac{1}{s_2^2}\right).
\eeq

\subsubsection{Entanglement in the presence of curvature} 
\label{sec.parallel_transport}

Consider creating an entangled pair, say of polarized photons or massive
spins,
which are
separated so far
that curvature and the nontriviality
of parallel transport becomes significant.  When considering an EPRB-type
experiment on two such separated spins, the question 
arises as to what direction the reference frame in the
neighborhood of one particle corresponds to in the neighborhood of another. Assuming there is curvature, parallel transport of the $z$-direction
from one spinning particle to the other is ambiguous since it is path dependent. Even though there may be a unique geodesic between the two EPRB measurements,
the common assumption is that one needs
to calculate the parallel transport of one spin's $z$-axis back (in time) along its worldline to where the pair
was created and then parallel transport that direction forward along the
other spin's worldline 
to the other spin. On one hand, if the curvature is known, this should allow one to predict how the reference frames of the EPRB measurements
ought to be aligned to meet the EPRB predictions. Experimental verification would constitute the measurement of an effect that involves essential parts of gravity and quantum theory, namely both curvature and entanglement.

Similarly to the special-relativistic effects of Section~\ref{ent-sr}, any
gravitomagnetic phase $\Delta\chi$ (c.f.\ Section~\ref{frame_dragging.sec}) which arises along the path of particles can be considered as arising from a local gate, and thus does not change the entanglement.
So long as the approximation of point particles that move on well-defined
trajectories holds, the parallel transport of the reference directions (for
massive particles) or local Newton gauge construction for photons will allow
preservation of the entanglement.

On the other hand, if such an experiment could be set up, it would allow one also to use a quantum effect to measure aspects of curvature.  To this end, one would perform an EPRB-type experiment with such entangled pairs, initially not knowing the curvature along the paths that the entangled particles took on their way from where they were generated to where they were measured. The aim would be to find and record that relative alignment of the local reference frames which optimizes the EPRB effect, i.e.\ which corresponds to the correct identification of the local reference frames. This would constitute a curvature measurement in the following sense. Assume that the curvature between the source of entangled particles and the EPRB measurement apparatuses changes. This could be detected in the EPRB experiments on subsequently produced entangled pairs, because for them the optimal alignment of the local $z$ axes would be different.

While the effects are tiny, and perhaps classical optical experiments may
have a greater chance of success in measuring properties of the gravitational
field, using entangled states to probe spacetime geometry
is interesting at least as a question of principle.
This proposal is also discussed in Ref.~\cite{KAJBPLZ03}.  There the
authors propose using NOON states in an interferometer to enhance the
observed signal.

In principle, EPRB experiments on entangled pairs of spins that travelled through regions of significant curvature need not only combine curvature with entanglement. They could also combine curvature and entanglement with quantum delocalization. This could be achieved if the entangled spins are made each to travel in a wave packet that is as large as the curvature scale. In this case, there is no unique path that the entangled particles take and therefore the rotation of their spins that arises from the curvature is quantum uncertain. Presumably, this still leads to an effectively fixed orientation between the two $z$ axes so that the full EPRB effect is observed. That is because as long as the particles do not significantly impact the spacetime, spacetime is not observing and not decohering the particles' paths.

This setup is likely to be exceedingly difficult to implement experimentally because of the difficultly of keeping track of individual entangled pairs over scales that are large enough so that the gravitational nontriviality of parallel transport becomes significant. However, intriguingly, this setup fundamentally involves both quantum and gravity effects without requiring high energies, much less energies close to the Planck scale. What makes this possible is of course that we are here not yet looking for quantum effects of gravity but merely for quantum effects that are influenced by gravity.

\subsection{The Fermi problem and spacelike entanglement tests }\label{Emm1}

The Fermi problem~\cite{Fermi} is a gedanken experiment put forward by Fermi as a causality check at a quantum level. The experiment consists of two spacelike separated atoms A and B at relative rest separated by a distance $R$. At some common proper time $t=0$, A is in an excited state while B is in the ground state, and the electromagnetic field is prepared in the vacuum state. The question that Fermi asked is ``Can the atom A decay to the ground state and provoke the excitation of the atom B at a time $t<R/c$?''

Due to intrinsic characteristics of quantum electrodynamics (namely the non-vanishing propagator of the electromagnetic interaction outside the light cone) the answer to this question has provoked a lengthy controversy~\cite{Ferm1,Ferm2,Ferm3} on the possible causal behaviour of the probability of excitation of the atom B.

Recently, a series of works demonstrated that a proper approach to this problem requires a
reformulation in terms of non-signaling and quantum non-locality~\cite{resin}. The atom B has a non-zero probability of getting excited outside the light cone, but this probability is completely independent of atom A. Therefore it cannot be used to carry superluminal information. The existence of nonlocal correlations outside the light cone cannot be used to violate causality.

Although some experiments to test this quantum foundational question have
been proposed in platforms such as superconducting circuits and optical
cavities, none of them has yet been carried out.

An experimental test of the Fermi problem would constitute a fundamental test of quantum field theory. The experimental difficulties of such an experiment come from the initial state preparation and the control of the process of switching on and off the effective interactions.

Being able to satisfactorily acknowledge this result in an experiment would require that we can guarantee that the detectors remain causally disconnected during the experiment while they are coupled to the same quantum field\footnote{The demands on the conditions of such an experiment could be relaxed requiring only that the interaction were disconnected before the spatial separation time ends and allowing the measurements to be carried out when causal contact is already possible. While much easier from the experimental point of view and still a proof of harvesting of vacuum correlations, this would constitute a weaker claim from the quantum foundations viewpoint, since causal influence on the measured correlations cannot be completely discarded.}. Let us call this difficulty the loophole of spacelike detection.

The Fermi problem is strongly related with the phenomenon first reported by Retzker, Reznik and Silman~\cite{reznik,CiracR} that two spacelike separated detectors can extract entanglement from the vacuum state of the field. In this scenario we have two spacelike separated detectors in the ground state at relative rest in a flat spacetime. These detectors are coupled to a massless field in a time dependent way that, prior to the activation of the interaction, is in the vacuum state.

At some point the interaction is switched on and the detectors interact with the field  during the short time they remain causally disconnected, then the interaction is switched off. The result is that the final state of the detectors presents quantum entanglement. Depending on the arrangement of spacelike separated detectors  the extracted entanglement can be very strong, even nearly reaching the maximally entangled state for the two detectors~\cite{Sabin2,Sabin2010,Sabin2012}.

The study of this problem connects with the fundamental results found in algebraic quantum field theory about the correlations in the vacuum state of the field between  local algebras of observables ``living'' in spacelike separated patches of the spacetime~\cite{Alegbra1,Alegbra2}. While detection of this phenomenon would be a test of the quantum non-locality in quantum field theory, and some proposals have been sketched~\cite{CiracR}, an experimental test 
is still to appear.

The amount of generated entanglement between two detectors at rest separated by a distance $R$ decays exponentially with the ratio $R/(cT)$ where $T$ is the interaction time. Namely, a lower bound for the entanglement of the two detectors' subsystem  after the interaction time $T$ is~\cite{reznik}
\begin{displaymath}
\mathcal{N}\sim \exp\left[\left(\frac{-R
}{cT}\right)^3\right]
\end{displaymath}
where $\mathcal{N}$ is the negativity of the two detectors' partial state~\cite{Negat}. For the detection of spacelike entanglement in a laboratory experiment, we again run into problems with the spacelike detection loophole: it is difficult to keep the relevant interaction switched on only during the time that the detectors are spacelike separated.

Indeed, a carefully controlled laboratory setting where we can have complete control on the interaction would have serious difficulties in carrying out the state preparation and state readout of the detectors within the time interval (of order of nanoseconds) that the detectors remain spacelike separated.

This spacelike separation loophole can be overcome using non Earth-bound
experimental platforms. The longer the distance between the two detectors, the longer we can keep the condition of spacelike separation. That means increasing the interaction time and therefore dramatically reducing the problems with the interaction switching and detector readout (see Table~\ref{tab.second}).

To ensure that we would be able to detect this entanglement extraction, the
detectors have to couple to the same field mode and coherence must be kept in
the quantum field mode probed. That means that any source of noise in the
relevant bandwidth should be kept to a minimum. The relevant bandwidth is given by the spectral response of the detectors. For instance, if the detectors are atom-based the spectral response of the detector is given by the gap of the atomic transition used.

As discussed in Ref.~\cite{reznik}, under these conditions, enough entanglement to violate  CHSH inequalities  can be effectively extracted for any separation distance $R$. The amount of entanglement that can be achieved between the detectors increases with the 
frequency gap, so an experiment using atoms whose energy difference between
the ground state and first excitation is of the order of visible light will
give better results than an experiment using detectors tuned to the microwave
or radio spectrum. This is beneficial for a possible experimental
implementation since isolating the mode probed by the detectors from noise in
the light spectrum is much easier than 
in the microwave or radio spectrum. To this regard, the requirement to reproduce the exact theoretical scenario in Ref.~\cite{reznik} is that the field mode probed is approximately in the vacuum state.

\begin{table}[ht]
\begin{center}
\begin{tabular}{|l|c|c|p{4.5cm}|}
\hline
\textbf{Kind of experiment} & \textbf{Typical distances $R$} & \parbox{3.75cm}{\textbf{Spacelike separation times $R/c \ge T
$}} & \textbf{Feasibility}\\
\hline
\hline
Tabletop&${\approx} 1$\,m & ${\approx} 3$\,ns & -- \\
\hline
Earthbound& ${\sim} 10$\,km & ${\approx} 30\,\mu$s & Possibly with superconducting qubits~\cite{Wallraff04}  \\
\hline
LEO satellite-based& ${\sim} 1000$\,km & ${\approx} 3$\,ms & Barely feasible with ions~\cite{Leibfried}\\
\hline
GEO satellite-based& ${\approx}$36,000\,km & ${\approx} 0.1$\,s &Well feasible with ions/atoms\\
\hline
Moon-Earth& ${\approx}$380,000\,km & ${\approx} 1$\,s & Feasible with macroscopic detectors\\
\hline
Solar System& ${\sim} 1$\,a.u. & ${\approx}\,500$\,s & Well feasible with macroscopic detectors\\
\hline
\end{tabular}
\caption{\label{tab.second} Interaction times for detectors at fixed relative distance in which the detectors only interact with the field while they remain spacelike separated.}
\end{center}
\end{table}

When going to longer separation distances we have to be very careful about this assumption on the mode with which the detectors interact: if the mode probed in this experiment is subjected to any kind of noise that makes invalid the assumption that the state of the field is approximately the vacuum, then the experiment will deviate from the hypothesis of the published theoretical works mentioned above. However, achieving noise insulation of a quantum channel in space does not seem extremely difficult.
In absence of direct solar illumination, the most important source of noise,
the CMB, can be avoided using visible light photons, which are widely detuned
from the peak of the CMB spectrum (a thermal spectrum at $T\approx 3$\,K).
Of course that would imply that the experiment should be screened from solar
radiation. Provided that the satellites are not receiving direct solar
illumination, the noise in the visible light channel can be, at least in
principle, kept to a minimum.
This suggests that the 
experiment could benefit from solar shielding of the spacelike separated detectors, with a design similar to what is planned for the James Webb Space Telescope~\cite{JWST}, or strategical placement in the shadow of planets and moons,  or maybe even on a hypothetical Lunar station.

The accomplishment of this kind of experiment would also have further
implications on tests of quantum field theory in non-inertial frames and in
the use of entanglement as a tool in cosmology, as we discuss 
in Section~\ref{EMM2}.

\subsection{COW experiments }\label{cow.sec}

In 1975 Colella, Overhauser, and Werner (COW) performed an experiment which
observed gravitational phase shift in a neutron beam interferometer~\cite{cow1}.  The underlying motivation for
this experiment (repeated with increasing precision over several years~\cite{cow2,cow3}) is to test the classical principle of equivalence in the
quantum limit. Despite many improvements to the original experiment, a small discrepancy of 0.6--0.8\% between theory and experiment remains~\cite{Kaiser2006Gravitat}.

For a beam of neutrons of mass $m$ and wavelength $\lambda$ entering an interferometer, the phase shift between the two sub-beams is
$$
\Delta \phi  = -  \lambda \frac{m^2 g}{2\pi \hbar^2} A \sin\alpha = - 2\pi \frac{ g}{\lambda v^2} A \sin\alpha
$$
where $A$ is the area enclosed by the trajectories of the two sub-beams, $\alpha$ is the tilt angle of the interferometer above the
horizontal plane and $g$ is the acceleration due to gravity.   The second equality follows from the de Broglie wavelength formula
(for neutron velocity $v$) and in this sense the COW experiment can be regarded as a gravitational redshift experiment for massive particles.  Upon replacing $v \to c$, this formula becomes identical to the phase shift formula for photons along different paths in
a constant gravitational field~\cite{cow4}.

\begin{figure}
\begin{center}
\includegraphics[height=6cm]{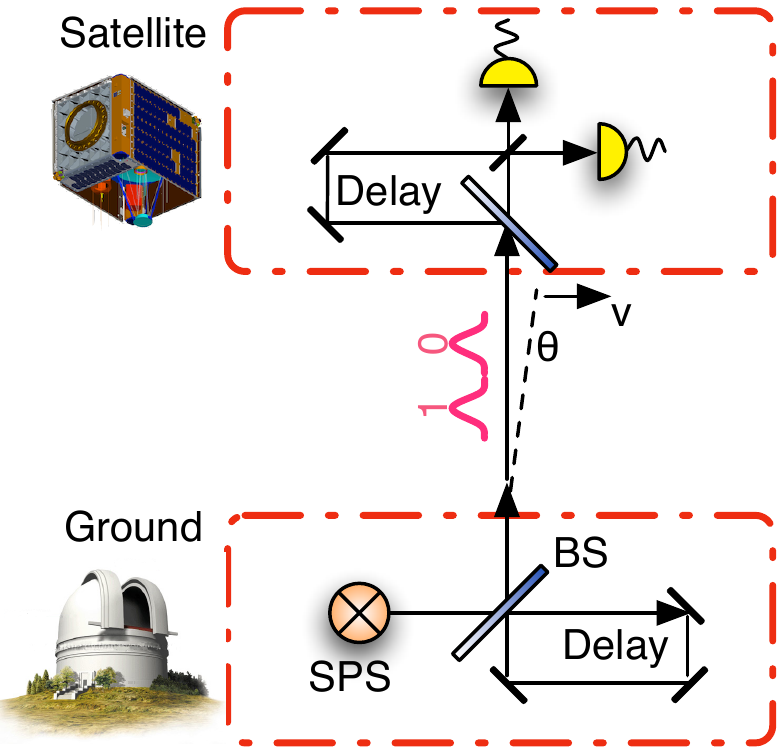}
\caption{\label{fig.cow} Optical COW experiment.
  The two delays could each be
  up to 20\,$\mu$s using $\sim$6\,km optical fibres, during which a LEO
  satellite will move about 15 centimeters.  The two beams differ by an angle
$\theta \sim$ 31 arc msec.}
  \end{center}
\end{figure}

It is therefore interesting to consider an optical version of the COW experiment using
transmission of a coherent beam to a satellite in LEO --- see
Figure~\ref{fig.cow}.  At the Earth's surface and in the moving satellite the beam is
coherently split by a semitransparent mirror, with one path going through
$l=6$\,km of optical fibre delay, while the other
path is directly transmitted.  The paths are recombined at the end to
construct an interferometer.  Since the upper fibre sits at a much higher
gravitational potential than the lower, it will experience a different phase
shift from the one on the ground, which will be picked up by the
interferometer.

Such an optical COW experiment could provide tests of gravitational redshift in the context of a quantum optics experiment. Specifically, in a weak gravitational field the redshift is given by the  difference of the Newtonian gravitational potentials $\varphi$ on the ground and at the satellite's orbit,
\beq
\Delta\omega=\frac{\Delta\varphi}{c^2}\omega\approx\frac{gh}{c^2}\omega,
\eeq
where $g$ is the free fall acceleration on the Earth's surface and $h$ is the
altitude of the satellite. The resulting phase difference between the two
paths is $\Delta\phi=\Delta k l$, where $k$ is the wave number. Hence, taking
the wavelength $\lambda = 800$\,nm and the altitude $h\sim400$\,km, one obtains quite a considerable phase difference:
\beq
\Delta\phi=\frac{2\pi l}{\lambda}\frac{gh}{c^2}\sim 2\, \mbox{rad}.
\eeq

Furthermore, due to the large speed of the satellite (ca.\ 8\,km/s) it will have moved by about 15\,cm (depending on the exact location of the satellite when the photons reach it) while the interferometer is closed, which effectively extends the satellite based delay. This has two consequences: first that the interferometer paths now cover an actual area and could be sensitive to rotational effects, and second, that in order to calibrate for zero-path-length, this delay, and consequently the satellite speed, must be very accurately measured, which could provide useful information for analysis of satellite motion.

Note that this optical COW  interferometry experiment would be carried out over much larger distances so
that the Earth's gravitational field is not constant.  The phase shift detected would then be general-relativistic as opposed to
one due to an accelerated frame (which the constant field near the Earth's
surface is equivalent to).   This would
constitute the first direct measurement of quantum interference due to curved spacetime. A detailed study of the visibility of interference fringes in the gravitational field can be found in Ref.~\cite{zcpb11, zcprb12}.

Higher order relativistic kinematic effects, similar to the Sagnac effect~\cite{sted03,mal00,cow-rev1}, especially for experiments with a pair of satellites, are also expected.

\section{Tests of quantum field theory in non-inertial frames}
\label{sec.qft}

The tests in Section~\ref{sec.entanglementSRGR} consider relativistic effects
on quantum mechanics and in special relativistic field theory.  Here we go further to explore how we might be able to probe the physics of quantum field theory in accelerating frames.  To our knowledge, no direct experimental test of physics in this regime has ever been performed.

\subsection{Test of the Unruh effect, entanglement fidelity and acceleration}
\label{unruh.sec}
An accelerated observer in the field vacuum as defined by an inertial observer would experience a thermal bath in a phenomenon analogous to the Hawking effect in a black hole.  This is called the Unruh effect~\cite{Unruh}. For an
acceleration $a$, the temperature of this thermal bath is
\begin{equation}
T = \frac{\hbar a}{2 \pi c K} \;,
\end{equation}
where $K$ is the Boltzmann constant.  At the surface of the Earth $a \approx
10$\,m/s$^2$, yielding $T \sim 4 \times 10^{-20}$\,K.

The first order effect of this thermalization is to act as a source of local noise in any mode of the field that may be probed by an accelerated  detector.  A more interesting effect is one in which (depending on the initial state of the field) a system initially prepared in an entangled state in an inertial frame may experience a variation in its degree of entanglement when measured by receivers that are
in relative acceleration. In the simplest scenario 
with a maximally entangled state of the field, it has been shown that entanglement degrades due to the presence of this Unruh noise~\cite{FM05,AM03}.

The magnitude of the acceleration necessary to observe these effects in, for example, an entangled state  of the electromagnetic field~\cite{Arbitrary}  in a field mode of frequency $\omega$ is
\begin{equation}
a\approx \omega c
\end{equation}
which for a field mode in the MHz spectrum would mean an acceleration of approximately $a\approx 10^{13} g$. This is already much smaller that the acceleration $a\approx 10^{22}g$ that one needs in order to observe a thermal bath warmer than the CMB.  However it is still too large to contemplate  direct observation in the gravitational field of any celestial body within the solar system. A satellite-based experiment aiming to look for the resulting loss of entanglement between an accelerating receiver on the Earth's surface and an inertial observer in orbit would be  unable to see this effect.

The preceding results were obtained using what is known as the ``single mode approximation'' (SMA)~\cite{Brusc}.  There has recently been rapid development in the study of quantum entanglement fidelity in non-inertial frames that goes beyond this approximation. For instance, it has been shown that beyond SMA and for some choices of the bipartite state, the accessible entanglement for an accelerated observer may behave in a non-monotonic way, conversely to the first results reported in Refs.~\cite{FM05,AM03}. This is due to inaccessible correlations in the initial states becoming accessible to the accelerated observer when his proper Fock basis changes as acceleration varies~\cite{MigC,CIvyty6}.

Furthermore, while the idealized quantum field modes used in previous studies
 are arguably impossible to reproduce in any experimental setting (due to their non-localized and highly oscillating spatial profiles~\cite{Brusc}), the study of the behaviour of  localized field states  has just recently started.  Several novel proposals have the edge on idealized scenarios for experiments where these effects can be detected.  This is the case of localized projective measurements~\cite{Drago1}, homodyne detection schemes~\cite{ralphhom} and accelerated cavities~\cite{CIvyty1,CIvyty2,CIvyty3,FLBL}.
In the latter case, there are very promising results regarding entanglement being generated instead of destroyed due to relativistic accelerations of optical cavities that can be, in principle, made detectable with relatively smaller accelerations~\cite{CIvyty3,CIvyty4,CIvyty5}.  However, concrete experimental proposals for the direct detection of these phenomena are still to be developed and it is not clear that with these settings we could take any advantage of the variations of the gravitational field strength in a solar-system based satellite experiment.

Nevertheless, an experimental test aiming to detect the Unruh effect can make use of the unique gravitational environment provided by a satellite platform. Some of the proposed experiments that have emerged from the field of relativistic quantum
information require high-precision 
measurements. An important case in point is  the Berry phase atomic
interferometry experiment, proposed in Ref.~\cite{BerryMM}, to detect the
Unruh effect.   This experiment exploits the fact that a  detector acquires a
geometrical phase due to its motion in spacetime. The phase differs for
inertial and accelerated particle detectors as a direct consequence of the
Unruh effect. This phase difference can be measured in an atomic
interferometry experiment in which one of the interferometer arms traverses a
potential fall that accelerates the atom.
The phase difference between the inertial and the accelerated atoms in one cycle of evolution is given by
\begin{equation}
\Delta\gamma_a=\arg\left(\cosh^2 q_a - e^{2\pi i G } \sinh^2 q_a\right)
\end{equation}
where $q_a=\arctan[\exp(-\pi\Omega_a c/a)]$ is a function of the atom's acceleration and $G$ depends on the atom properties (energy between the ground and the excited state, coupling strength, etc.) as detailed in Ref.~\cite{BerryMM}.

This method is sensitive to accelerations 
$10^9$ times smaller than other suggested attempts to probe this effect~\cite{ChenTaj,Crispino}. Unfortunately, the necessary precision in the relative phase between these detectors is so high ($\Delta\gamma_a \approx 10^{-3}$ to $10^{-5}$) that Earth's gravity can interfere with the reading of the atomic interferometer required for detection~\cite{Atomin}: the interferometer could be sensitive to variations in the gravitational field of order ${\sim} 10^{-15}g$~\cite{Atomin2}.  However, the freely-falling environment of the satellite would significantly reduce such interference, providing a great advantage for a realistic experimental proposal.
While this experiment does not directly probe physics at large length scales
by performing inter-satellite quantum communication, we deem it important for
(1) its apparent need for a free-falling space environment, and (2) its
attempt to directly explore the effect of acceleration on quantum entanglement.

\subsection{Gravitationally induced entanglement decorrelation} 
\label{sec.Ralph}

The standard description of quantum fields in curved spacetime~\cite{Birrell}
allows quantum entanglement to survive unchecked in a wide variety of
gravitational backgrounds. In principle, then, entangled photons created
locally and transmitted to regions of different gravitational potential will
still possess all the entanglement they started with, and local detectors can
be suitably designed that will reveal this.\footnote{These would need to
  take into account time-of-arrival delay, redshift, polarization, mode shape
  modification, phase locking, etc.} This would be challenging, of course,
but such designs are at the heart of most every proposal in this article.

An alternative formulation of the behaviour of photons in curved spacetime was proposed by Ralph in a series of papers~\cite{Ral05,Ral06,Ral07,RMD07,RMD09} originally designed to deal with the hypothetical question of quantum information propagation through closed timelike curves but later revealed to predict measurable decoherence induced by gravitation. The essence of the proposal is to supplement ordinary field theory with an additional degree of freedom called an \emph{event operator}, which is associated with the detectors used to measure the field quanta in a given setup. In flat spacetime with detectors in the same reference frame, the event operator is of no consequence because time passes identically for all detectors. But when two detectors are in regions of differing gravitational potential, their local clocks run at different rates and fall out of sync, little by little. When the amount by which they desynchronize during the photons' times of flight is longer than the timing resolution of the detector, then, although the delay in the time of arrival can be accounted for in the detector design, the presence of the event operator (whose duration is associated to the detector's temporal resolution) ensures that the two photons lose coherence.\footnote{Since this is a non-standard extension to quantum field theory in curved spacetime, we direct the reader to the literature for further details~\cite{Ral05,Ral06,Ral07,RMD07,RMD09}.}

\begin{figure}[Htbp]
\begin{center}
\includegraphics[width=0.41\columnwidth]{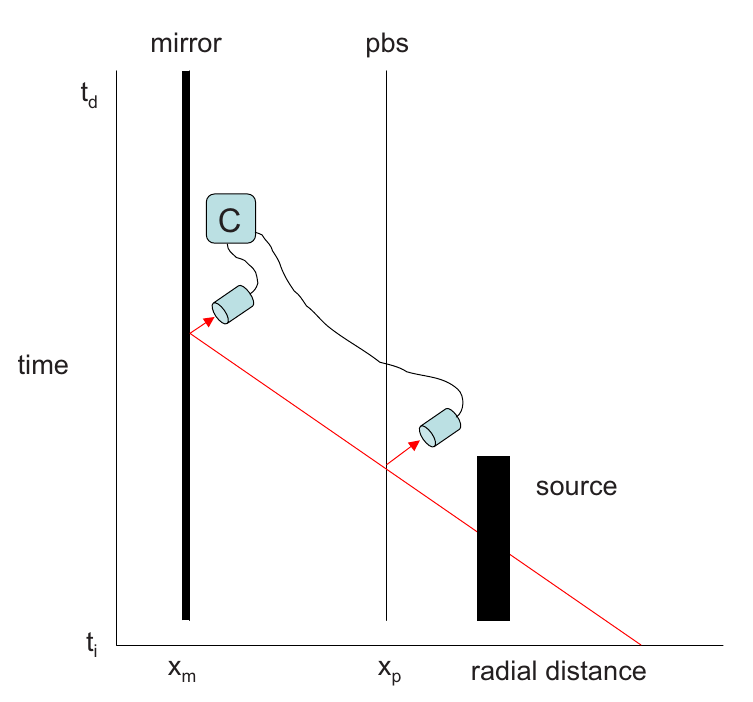}
\includegraphics[width=0.41\columnwidth]{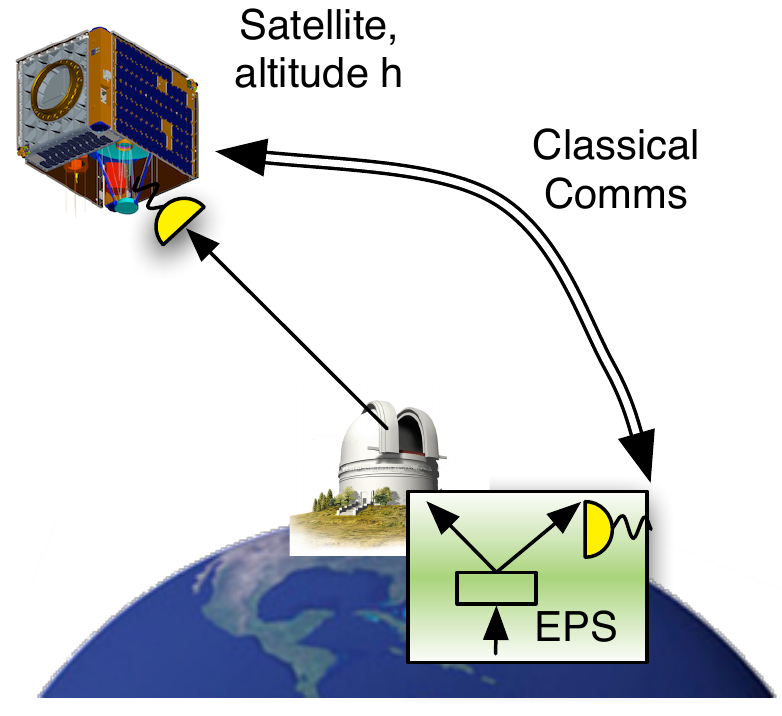}
\end{center}
\caption{Left: The original proposed scheme to test gravitationally induced decorrelation. A source prepares a pair of orthogonal polarization
modes with entangled photons. Two modes initially propagate towards
the surface of the Earth. Using the polarizing beamsplitter (pbs), two modes
are separated at height $x_{p}$. The detectors located at height
$x_{p}$ and $x_{d}$ pick up the signals and feed the measurement
result to the correlator via classical channels. Figure taken from Ref.~\cite{RMD09}. Right: A possible  implementation could employ an uplink of the quantum signals from the source on the ground to the receiver in the satellite, which inverts the signs of the gravitational potential difference, but should lead to the same effect. The entangled photons prepared via SPDC are initially perfectly correlated and spatially
degenerate. Two single photon detectors record the detection times
of photons and stream the data to a computer, and they are sent from the satellite to the ground for analysis. \label{fig.ralph_setup}}
\end{figure}

The scheme proposed in Ref.~\cite{RMD09} involves preparing a pair of entangled photons via spontaneous
parametric down-conversion (SPDC). One photon is measured directly on the
ground station after a time delay, while the other is sent to the satellite, traversing a
non-uniform gravitational field. Figure~\ref{fig.ralph_setup}
illustrates the process.

Using the localized event operator introduced in the theory, the maximum coincidence detection rate of two photons should
decline due to intrinsic decoherence by the curved spacetime.  The difference
in proper time~$\Delta$ between the two detectors during the time of flight
of the photons sets the timescale for the event operators in question, and
therefore sets the timescale on which decoherence will take place. When this is larger than the detector temporal resolution~$d_t$, decoherence occurs. In the specific model proposed in Ref.~\cite{RMD09}, the reduced correlation function is given by
\begin{equation}
\label{eq:Ralphcorrelation}
 C=C_\mathrm{max} \exp\left(-\frac{\Delta^{2}}{4d_{t}^{2}}\right),
\end{equation}
where $C_\mathrm{max}$ is the
correlation function we should observe in flat spacetime, and $d_{t}$ is the
temporal resolution of the photon detectors.

\begin{figure}[Htbp]
\begin{center}
\includegraphics{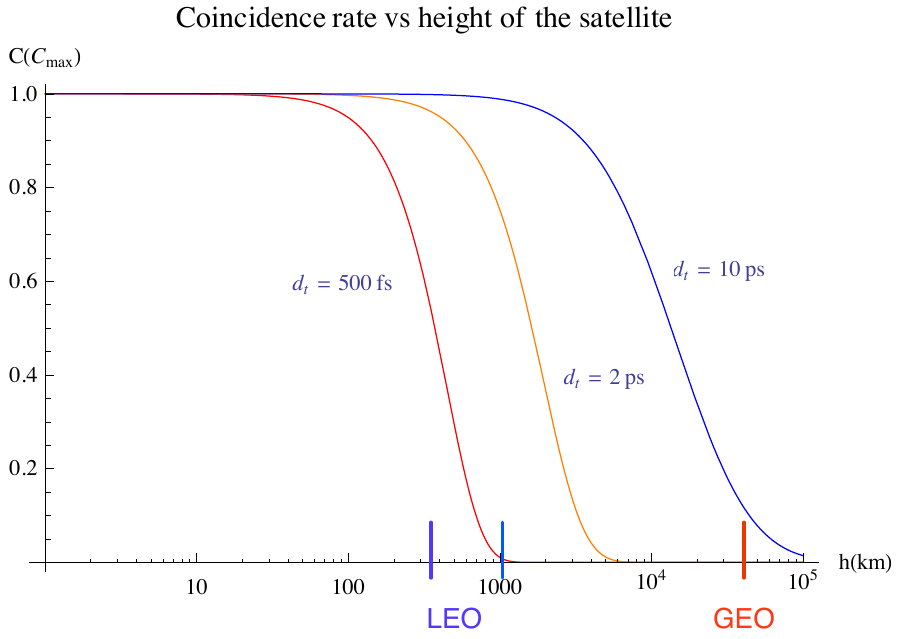}
\end{center}
\caption{Coincidence predictions from the experiment proposed in Ref.~\cite{RMD09}.  The coincidence detection rate
(i.e.\ detection events in both detectors) as a function of temporal
difference of the detection $t_{d2}-t_{d1}$ should be peaked around
the light travelling time difference in the two ``arms''. The maximum coincidence
rate $C$ describes the photon correlation upon detection, within the (intrinsic) detection time $d_t$. \label{fig.ralph_coincidence}}
\end{figure}

If the photodetector response time is 500\,fs, then the decoherence
should easily be observed for satellite altitudes of 400\,km,\footnote{This is a conservative estimate compared to the 200\,fs quoted in Ref.~\cite{RMD09}, which would result in decoherence above 90\,km.}
as illustrated in Figure~\ref{fig.ralph_coincidence}.
With a GEO satellite (36,000\,km), the effect would even be stronger, and even
with 10\,ps time resolution (the timing resolution of a typical contemporary photodetector), a strong decorrelation effect could occur.

Due to the careful timing required, this scheme is challenging but possibly doable. It is interesting to note that the predicted decoherence induced by gravity applies only to quantum entanglement, and is in addition to any spreading of classical correlations.
Because there are many sources of decoherence for photons travelling between satellites and the ground, it will be much easier to refute
the event-operator hypothesis than to confirm it ---
if we were able to generate entanglement beyond what is predicted by Eq.~(\ref{eq:Ralphcorrelation}), the hypothesis as proposed would be refuted. In contrast, if decoherence were found, one could (for example) employ a satellite in an elliptical orbit to perform the test
at different heights to see whether 
the decorrelation varies in accord with Eq.~(\ref{eq:Ralphcorrelation}).

We also note that an alternative to the setup of Figure~\ref{fig.ralph_setup} (right panel) is to place a retroreflecting mirror on the satellite instead of a detector. This would require two passes through the atmosphere instead of just one (causing additional loss and distortion) but would have the advantage of cutting in half the height at which an effect can be seen. Thus, the theory would predict the same curves as in Figure~\ref{fig.ralph_coincidence} but with each tick mark on the horizontal axis replaced with half its current value. Depending on the details of the experiment, this might be advantageous.

\subsection{Probe the spacetime structure by spacelike entanglement extraction} 
\label{EMM2}

It is known that the expansion of the Universe can produce particle creation when conformal symmetry is broken~\cite{Birrell}. It was proven, for example, that in an FRW universe with 
metric tensor 
of the form
$ds^2 = dt^2 - a(t)^2 dE^2$ ($dE^2$ being the line element of flat 3-space)
for massive fields or for massless fields not conformally coupled to the curvature, entanglement is generated in quantum fields during periods of fast expansion such as inflation~\cite{caball,fermxpanding}.

When conformal symmetry is present, it is also well known that comoving
detectors (i.e.\ detectors that see isotropic expansion of the universe, such
as --- to an extremely good degree of approximation --- detectors placed in satellites) respond by clicking, even in the conformal vacuum, due to the difference between their proper times and the conformal time~\cite{Birrell}. Gibbons and Hawking~\cite{GibHawking} showed that a comoving detector in a DeSitter background (exponentially expanding universe, $a(t)=e^{\kappa t}$) has a thermal response to the conformal vacuum state of the field whose temperature is related to the presence of a cosmological horizon and is proportional to the Hubble parameter.

However, a single comoving detector would constitute a poor probe for the
structure of the spacetime. For instance, in the scenario mentioned above,
particle detection would be not be enough to distinguish an exponentially expanding universe from a thermal source in a flat spacetime. On the other hand, using entanglement induced in spacelike separated detectors~\cite{Nick}, a way to distinguish the conformal vacuum of a scalar field in a DeSitter universe from a thermal bath in flat spacetimes was proposed: while a detector sees the same thermal bath in both scenarios, the way in which two separated detectors become entangled senses a difference.

This is just an example of a more general phenomenon still to be explored: quantum correlations acquired by spacelike separated detectors interacting with the same quantum field~\cite{reznik} are dramatically sensitive to the spacetime background and the state of motion of the detectors. For instance, it has been argued that due to general relativistic effects, two counter-accelerating detectors can extract more entanglement from the vacuum state than inertial ones~\cite{Drago1}.

It is important to keep in mind that no quantitative results in realistic
regimes has been obtained yet: to reproduce the scenario depicted
in Ref.~\cite{Nick}, two satellites should be sufficiently separated in comoving
distance so as to be separated by the cosmic horizon (so that the expansion of the universe will prevent them from mutually communicating in the future while they remain able to communicate with the home planet which is equidistant from them), something beyond the reach of any thinkable experiment with current technology. However, the effect of the spacetime structure on spacelike entanglement is currently starting to be studied and results will be arguably produced soon for detector separations of the order of the solar system and smaller scales. These results will tell us how much sensitivity such a hypothetical experiment should have.

While potentially useful as tools to probe the geometry of the spacetime provided our ability to measure those correlations, all these experiments are subjected to the same spacelike separation loopholes as those described in Section~\ref{Emm1}. Having arrays of two or  more detectors in space will allow longer-lasting spacelike separations. Then, interaction times can be made longer while preserving the spacelike separation condition.

\section{Quantum gravity experiments}\label{sec:q_grav_expts}

\subsection{Background}

The direct observation of quantum gravity effects,
i.e., of quantum fluctuations of curvature, is very difficult,
the reason being that these are expected to be significant
near the Planck scale
($10^{-34}$\,m, assuming $3+1$ dimensional spacetime) which is well over
a dozen orders of magnitude beyond the 
scales that particle
accelerators can probe. So far, the best potential for 
experimental
evidence for 
quantum gravitational effects 
stems from 
cosmology. The standard model of cosmology
holds that the observed inhomogeneities
in the 
CMB were seeded by 
joint quantum fluctuations of a scalar field called the inflaton
and the scalar-derived part of the metric. The theory of cosmic inflation is
experimentally very successful but, due to coordinate gauge dependence, one
cannot strictly distinguish the observation of temperature inhomogeneities in
the CMB that are due to quantum fluctuations of the metric from temperature
inhomogeneities that arose from quantum fluctuations of the inflaton
field. In the future, if experimental efforts to detect the curl component of
the polarization field of the CMB 
using the PLANCK satellite succeed they could provide a major step
forward. This is because cosmic inflation holds that this curl field
originated in a clean quantum gravity effect, namely in quantum fluctuations
of the tensor-derived part of the metric during the very early inflationary
phase of the universe. In the longer term it is indeed conceivable that a
quantum gravitational effect could be observed in the CMB polarization
spectrum. This is because the quantum fluctuations that ultimately caused the
temperature fluctuations in the CMB were frozen in magnitude when their
wavelength was that of the Hubble length during inflation.

Several independent arguments suggest that the Planck-scale structure
of spacetime should not be describable as an ordinary smooth pseudo-Riemannian geometry,
as a result of the interplay between quantum theory and general relativity,
both of which possess physical effects that would be non-negligible at the Planck scale.
Several alternative models for the description of the Planck-scale structure
of spacetime have been developed over the years, and a detection of some manifestation of such
an underlying alternate spacetime structure would have enormous scientific impact, in that it
would be an important clue toward
   the construction of a quantum theory of gravity. 
The correct prediction of the order of magnitude of the 
 cosmic acceleration~\cite{cosmic_acceleration,SorkinLambda} may be interpreted as initial observational
  evidence that spacetime possesses some non-smooth structure around the
  Planck scale.

Successful paradigms for the exploration of these issues for the
Planck-scale structure of spacetime are 
``spacetime discreteness'', ``spacetime noncommutativity'', and a
third, which 
posits that spacetime is simultaneously continuous and discrete, in the
same mathematical way that information can have this character~\cite{Kempf}.

A theoretical signature of the third type of 
ultraviolet cutoff 
is the deformation of the uncertainty principle so that it yields a finite
lower bound on position uncertainties~\cite{uncertainty_reln}. Experimental
signatures have been worked out and much discussed in the context of the
predictions of  inflationary cosmology for the cosmic microwave background~\cite{Kempf2}.
In particular, this quantum gravity cutoff would impact both the CMB
temperature spectrum as well as its polarization spectrum. The scale of the
predicted quantum gravity effects is optimistically at the level of $\sigma =$
(Planck scale) / (Hubble scale during inflation) $\approx 10^{-6}$, though
some studies suggest that the effect may only be as large as $\sigma^2$.
Detection of such an effect, however, falls under the category of
observational astronomy, and thus is not within 
the purview of active
tests which we consider in this paper.

As we shall discuss later in this section, one may look for several
possible direct or indirect manifestations of the microscopic structure of spacetime.
Among the indirect manifestations, much attention has been devoted in the
literature to three possible outcomes for the fate of the Lorentz
symmetry of special relativity: (1) Lorentz symmetry may well be preserved
near the Planck scale; (2) Lorentz symmetry may be ``broken'' near the Planck
scale, with associated emergence of a preferred (``ether'') frame; or (3)
one might have the so-called doubly-special-relativity scenario, with a
``deformation'' of Lorentz symmetry providing for Planck-scale modifications of
the laws of transformation among inertial observers, but preserving the
relativity of inertial frames.

For scenarios
based on spacetime noncommutativity, 
the analysis of observable spacetime features is still at an early stage.
Until very recently, spacetime noncommutativity was studied mainly indirectly
through its implications for momentum-space properties, because the spacetime
features appeared to be puzzling. Only very recently~\cite{smolin} was it understood
that these spacetime features were to be described in terms of relative
locality (see later parts of this section). We expect that over the next few
years this should allow a sharper handle on the spacetime aspects of the
phenomenology of spacetime-noncommutativity models.

\subsection{Lorentz invariant diffusion of polarization from spacetime
  discreteness}\label{diffusion.sec}

With regard to the 
paradigm of spacetime discreteness, an important question 
is whether one expects the discrete
structure to violate the Lorentz symmetry. 
Regular
lattices, such as salt crystals, do violate Lorentz invariance, however
``random lattices'', analogous to the arrangement of molecules in a water
droplet, do not.\footnote{This ``Euclidean signature analogy'' may appear
  oversimplified, in that random lattices can also be Lorentz violating.
  From the perspective of the Lorentzian signature, an important point is
  that Lorentz invariant lattices consist of nodes which have an infinite
  number of nearest neighbors, while lattices of finite valence inevitably
  violate the Lorentz symmetry.}
Work aimed at exploring phenomenological consequences of Lorentz violating
discreteness
has been advancing already for about a decade, however relatively little is
known about potentially observable consequences of Lorentz invariant discreteness.

If one were to imagine particle propagation on a discrete background, one might
consider a simple model in which the particle jumps from one lattice site to
the next, in some fashion.  For a random lattice, the jumping will be
described by a random process.  In the limit of an infinitely dense lattice,
such random jumping leads to a diffusion process.\footnote{This assumes ``local
  jumps'', which converge to a continuous path in the infinite density limit.
  More general processes are possible, in which one considers limits of paths
  with are not continuous.} 
Thus one may expect spacetime discreteness to manifest itself in terms of a 
diffusion process for particle propagation.

\subsubsection{Lorentz invariant diffusion of CMB polarization}

Given the existence of rather tight bounds on Lorentz-symmetry violation,
one is motivated to
look for evidence of such a diffusion process
which respects Lorentz invariance.
Contaldi, Dowker, and Philpott have considered the potential effect
on the polarization of light which propagates over long distances~\cite{CDP10}.
They write down the most general diffusion process on the polarization state
space for a photon, the Bloch sphere $\mathcal{B}$:
\begin{equation}
\frac{\partial \rho}{\partial \lambda} = \partial_A \left( K^{AB}
  n \partial_B \left( \frac{\rho}{n} \right) - u^A \rho \right) \;.
\label{eqn.tensors}
\end{equation}
Here $\rho$ is a probability density on $\mathcal{B}$, $\lambda$ is affine
time, and $n$ is a fixed ``density of states'' on $\mathcal{B}$.
$K^{AB}$, a symmetric positive semidefinite 2-tensor, and $u^A$, a
vector, are phenomenological parameters of the model.  They are constrained
by the model only insofar as they respect Lorentz invariance.

It can be shown that Lorentz transformations act as polar
rotations on $\mathcal{B}$~\cite{CDP10} (c.f.\ Section~\ref{lorentz-polar}).
Thus any tensor fields $K^{AB}$ and $u^A$ which are functions of only the polar
angle give a Lorentz invariant diffusion model.  In order to look for a
testable prediction, the authors restrict attention to linear polarizations,
which lie on the equator of $\mathcal{B}$.  Then Lorentz transformations act
as rotations around the equator, and all that remains are two parameters, $c$
and $d$, which govern the rate of diffusion and drift respectively, by
\begin{equation}
\frac{\partial \rho}{\partial \lambda} = c \frac{\partial^2 \rho}{\partial
  \beta^2} - d \frac{\partial \rho}{\partial \beta} \;,
\label{eqn.cd}
\end{equation}
where $\beta$ is an azimuthal coordinate. 
The affine parameter $\lambda$ can be given a Lorentz invariant meaning as
the ratio $t/(h \nu)$, where $t$ is a time coordinate in a frame for
which the photon frequency is $\nu$, and $h$ is Planck's constant.  Then
Eq.~(\ref{eqn.cd}) can be written as
\begin{equation}
\frac{\partial \rho}{\partial t} = \frac{c}{\nu} \frac{\partial^2 \rho}{\partial
  \beta^2} - \frac{d}{\nu} \frac{\partial \rho}{\partial \beta} \;.
\label{eqn.cdt}
\end{equation}

The authors
place tight bounds on the observed drift and diffusion of polarization
from observations of CMB polarization. 
They calculate that, for
\begin{equation}
\Phi = \arctan\left(\frac{U}{Q}\right) \to \Phi' = \Phi + \chi
\end{equation}
and
\begin{equation}
\mathcal{P} = \sqrt{U^2 + Q^2} \to \mathcal{P'} = e^{-\mu} \mathcal{P} \;,
\end{equation}
with Stokes parameters $Q$, $U$, and $V$, $\chi \ltsim 0.1$ and $\mu \ltsim
0.025$.
For a constant frequency $\nu$, one can integrate Eq.~(\ref{eqn.cdt}) to get
$\chi = td/\nu$ and $\mu = 4tc/\nu$.
Using this, one can compute bounds on the parameters of Eq.~(\ref{eqn.cd}) of
$d \ltsim 4 \times 10^{-8}\,\mathrm{s}^{-2}$ and
$c \ltsim 2 \times 10^{-9}\,\mathrm{s}^{-2}$.

\subsubsection{Testing for spacetime discreteness with satellites}

In the context of a satellite
experiment involving transmission of optical photons to LEO,
these parameters should lead to measured values of $\chi \ltsim 12 \times 10^{-26}$
and $\mu \ltsim 3 \times 10^{-26}$.

No observational bounds yet exist for circular polarization.  A simple
experiment to create such a bound could consist of preparation and detection
of circularly polarized photons that traverse a large distance.  For our
satellite setup, one could prepare a weak coherent pulse or individual
photons into a known circular polarization state, and perform state
tomography on the receiving end, to reconstruct the polarization state of the
received photons.  Such an experiment would then set bounds on the values of
the tensors in Eq.~(\ref{eqn.tensors}).  It is not entirely clear whether one should
expect a stronger signal from the greater statistics afforded by a large
photon number, because individual photons may experience greater diffusion
than larger a wave pulse.  Thus the experiment could tune laser power to find
the largest signal.

\subsection{Decoherence and spacetime noncommutativity} 
\label{noncommutativity.sec}

The alternative of studies of spacetime noncommutativity finds motivation in the
study of the 3D-quantum-gravity toy model.\footnote{3D quantum gravity
does share a few of the conceptual challenges of ``real'' 4D quantum gravity, however it is far simpler.}
Recent results
establish that for 3D quantum gravity (exploiting its much greater simplicity than the 4D case)
one is able to
integrate out gravity, reabsorbing its effects into novel properties
for a gravity-free propagation of particles.  The resulting noncommutative spacetime
picture is affected by two main features:
\begin{enumerate}
\item the spacetime symmetries are described by a Hopf algebra (rather than admitting the
standard Lie-algebra description) and in particular Lorentz symmetry is deformed;
\item the fuzziness of worldlines receives a Planck-scale contribution (in addition to
the one of ordinary quantum-mechanical origin).
\end{enumerate}
Thus the spacetime-noncommutativity programme can motivate searches for departures from
classical (Lie-algebra-described) Lorentz symmetry, and searches for effects linked with
the additional source of worldline fuzziness, such as minute losses of quantum-mechanical
coherence.

The setup with a weak coherent pulse of
photons discussed in the previous subsection
is also of interest from the perspective of spacetime noncommutativity.
As mentioned, the Planck-scale-induced fuzziness of worldlines is expected
to effectively produce loss of quantum-mechanical coherence, which should
grow over macroscopic distances: on table-top experiments the effects
would be completely negligible because of the Planck-scale suppression,
but there would be significant interest in bounds established
using the large distances involved in a satellite setup.

It is likely that definite predictions, which therefore could be used to set
bounds on the relevant parameters, will mature over the next two or three years,
taking as starting point preliminary results such as those in
Ref.~\cite{FreidelLivine06,Arzano2007}.

\subsection{``Relativity of Locality'' from doubly special relativity} 
\label{DSR.sec}

Deformations of Lorentz symmetry of the type 
motivated 
by spacetime-noncommutativity 
generate interest 
in the recent proposal of the ``relative-locality
framework''~\cite{smolin}.
In this proposal one endows momentum space with the role of ``primitive''
entity for the definition of physical observables, with spacetime only introduced
as a derived entity.  The geometry of momentum space, assumed to be non-trivial
(i.e.\ different from the trivial flat/Minkowskian geometry usually assumed
for momentum space), takes center stage, with the Planck scale playing the role
of scale of curvature of momentum space.

Several features of the geometry of momentum space could be of interest for setups
suitable for satellite operation. A particularly interesting setup from this perspective
is the one of interferometers studying phase differences produced through splitting
 beams  in energy space
 (see, e.g., Ref.~\cite{Amelino-Camelia2004}).
 In such setups,
a monochromatic wave with frequency $\omega$  goes through two or more
frequency doublers (or even ``comb'' generators) so that in the end one can perform
interference studies on beams (thereby produced by splitting the original beam in energy space)
which follow the same path in spacetime, but different paths in energy-momentum space.
If the geometry of momentum space is indeed nontrivial, with Planck-scale curvature,
such setups should produce testable features. The features would once again be tiny for table-top
experiments, but potentially observable for satellite setups.

\section{Quantum communication and cryptographic schemes}
\label{sec.qcommunication}

Quantum information allows one to perform communication tasks that are beyond the possibilities of classical systems. One such example is quantum cryptography, or more correctly titled quantum key distribution (QKD),
 in which secure keys are established using non-orthogonality of quantum
 mechanical signal states~\cite{RevModPhys.74.145,scarani-2009-81}.  Such states can be easily created using optical signals. The non-orthogonality of the signals ensures that no eavesdropping can be performed without introducing changes to the signals.

Performing such quantum protocols with satellites is an important
application, because it can allow for expansion of a potential service to global scales.
When such quantum applications are implemented in Space,
all the physical phenomena that affect the photons and their quantum states must be
understood and accounted for
such that one can reliably model the effective quantum channel to achieve the desired performance.

\subsection{Quantum cryptography with satellites }
Many proposals for QKD using satellites have been studied theoretically and with proof-of-principle experiments, e.g.\ Refs.~\cite{1367-2630-11-4-045017,kurtsiefer02,KAJBPLZ03,Buttler00a,VJTABUPLBZB08,UTSWSLBJPTOFMRSBWZ07}. We  therefore refer the reader to these references for more details.

The extensive analysis of the quantum link performance provided in those references shows that it should be feasible to exchange several millions of secure bits per month with a quantum satellite located in LEO. Using a satellite in GEO is very interesting too as its stationary location allows better accessibility of the system. Another important scenario is to send two entangled photons from the satellite towards two ground stations, enabling them to use the quantum correlations  directly to generate secure keys on the ground without having to trust the security of  the satellite. An entangled photon source in space could achieve useful key rates (several kb per month) depending on the technology that is implemented, and is currently envisioned in the Space-QUEST project being pursued in Europe~\cite{Armengol2008}.

\subsection{Quantum tagging}\label{tagging.sec}
Many of the most interesting recent developments in quantum
cryptography make use of relativistic signalling constraints
as well as the properties of quantum information to ensure
security.   Typically, these schemes become more efficient
and easier to implement when the separations between secure
sites become large, so that the delays caused by generating
and processing signals are small compared to the separations.
In particular, space-based implementations are more efficient
than terrestrial implementations.

Satellite-based implementations are also particularly natural
for the recently proposed protocols for \emph{quantum tagging} (also
called quantum position authentication).  The aim of these
protocols is to guarantee the location of a valuable (or
perhaps dangerous) object or person by exchanging signals with
distant secure sites.   For example, one might want to establish
a secure satellite network to verify the location of objects
or people on Earth, taking advantage both of the security
offered by satellites (which cannot easily be hijacked or compromised,
and especially not without detection) and of the large area of
Earth in their direct line of sight.

Quantum tagging protocols~\cite{patent,KMS10,Kent-tagging2} (see
also Refs.~\cite{Buhrmanetal, Malaney} for independent proposals)
aim to use the properties of quantum information and the impossibility of
superluminal signalling to guarantee the
location of an object that is distant from trusted secure sites.
These protocols are designed to deal with an adversarial scenario,
in which an enemy may be trying to relocate the object while fooling
the tracker by sending fake messages and images.  (For example, a
tagged prisoner wishing to escape might spoof
his tagging device so as to persuade the authorities that he remains in
custody.)

The overall level of security attainable depends on the properties of
the tag~\cite{Kent-tagging2}.   In one interesting and practical
scenario, the tag is assumed able to contain secret data.   This allows
perfectly secure tagging in any region within the convex interior
of the set defined by the secure signalling sites --- within a tetrahedron,
for example, for four non-coplanar sites.
The resources required are (only) a QKD link
between one site and the tagged object, together with (approximately)
light speed classical communication between the object and all four
sites.   Indeed, a technological proof of principle could be achieved
using a single satellite and two ground stations, with a QKD link
between the satellite and one ground station, together with
light speed communications (for example line of sight radio or
light pulse signals) between the satellite and both stations.

\subsection{Quantum teleportation with satellites}
\label{teleportation.sec}

Quantum teleportation allows the transfer of an arbitrary and unknown quantum state from one location to another through the use of an entangled pair resource~\cite{BBCJPW93}. By doing a joint Bell state measurement on the quantum state and one half of an entangled pair and sending the measurement results to a user with the other half of the entangled pair, the distant user can can recover the original quantum state. The key operation for this protocol is the Bell state measurement, which imprints the quantum information of the original photon onto the entangled photon pair, one half of which is transmitted to the receiver. Standard quantum theory places no bound on the distance at which teleportation may be accomplished. This process is highly interesting from a fundamental viewpoint, but also a crucial concept for interlinking  quantum computers using quantum networks.

Experiments have been performed on the ground up to a distance of
143\,km~\cite{Jin2010,Ma:2012fk,Yin:2012uq}, and in principle could be extended further.  However, as with other experiments, planetary extents limit the feasibly achievable teleportation distance that can be tested. Moving to a satellite platform as either source or receiver (or both) significantly raises these limits.

In a downlink scenario, a source of entangled photons would be placed on the
satellite with one half of the entangled pair being sent to a ground station
over the free-space link. The ground station would have its own photon source
to generate the quantum state which is to be teleported. The ground station
would interfere the received photon from the entangled pair with the photon
meant to be teleported in a joint Bell state measurement. The results of the
measurement would then be sent over the classical communication channel to
the satellite. There are then two options on the satellite, either
\begin{enumerate}
 \item the second photon from the pair generated on the satellite could be stored in a quantum memory (in this case perhaps a polarization stabilized optical fibre) until the classical measurement results were received allowing the correct feed-forward operation to be applied in all cases to recover the teleported quantum state, or
 \item no quantum memory would be used and the data would be post-selected onto those cases where the classical measurement results indicated that no correction operation needed to be performed.
\end{enumerate}
The satellite would also require the necessary equipment to analyze the final state of the teleported photon to verify that the teleportation protocol succeeded.

In an uplink scenario, the entangled photon pair source would remain on the ground while the satellite would contain a receiver system as well as an extra photon source capable of creating the quantum state which was to be teleported. The satellite would receive one half of the photon pair via the free-space link and perform a joint Bell state measurement on it along with a photon from the extra quantum source. The results of the measurement would then be classically communicated to the ground station. Lastly, the ground station would analyze the state of the photon with quantum state tomography to ascertain that the photon state was in fact successfully teleported. The advantage of this scenario is that there would be far fewer limitations on storing the photon that remained at the ground station in a quantum memory and it would be possible to utilize the latest technology. This would mean that it would be much more likely to perform an experiment that allowed the ground station to perform the necessary feed-forward correction in order to recover the teleported quantum state in all cases rather than post-selecting only those cases for which no correction was necessary. The obvious disadvantage is the practical complication of having apparatuses to perform a Bell state measurement on the satellite.

We may further consider an entanglement swapping scenario, where a satellite could receive two photons from independent ground stations and perform the Bell state measurement, entangling the remaining photons between the two stations. This requires quantum memories such that the photons on the ground can be stored until the Bell state measurement is performed.
 Such a scenario would be an important step towards the long-term goal of realizing global quantum repeater networks. The satellite would serve to establish an elementary link between two ground stations separated by a large distance. Several such links could then be connected by entanglement swapping operations between memories on the ground in order to establish entanglement over global distances on earth. Such a satellite-based approach to quantum repeaters would be significantly more powerful than an approach based purely on optical fibres on the ground, as reviewed recently in Ref.~\cite{Sangouard2011}.

Note that, in all the proposed experiments, the Bell state measurement
would involve at least one photon that has travelled over a long
distance (from the satellite to the ground or from the ground to the
satellite respectively). This is qualitatively different from
recent ground-based long-distance teleportation experiments
\cite{Jin2010}, where entanglement creation and Bell state
measurement were performed at the same location.

The applications of quantum teleportation between a satellite and the Earth's surface will most likely become very relevant once quantum information processors are widely deployed and need to be interconnected. In the intervening time, it will have a very important motivational impact for generating interest in this research, perhaps even more so than its direct impact on our understanding of fundamental physics. The importance of quantum teleportation experiments could therefore extend beyond its value as a direct test of physical theory or in providing a
useful service.

\section{Techniques which can be used to gain accuracy or isolate certain effects}
\label{sec.spekkens}

The ideas mentioned in this section can be thought of more as techniques
which may be useful for achieving the scientific aims of some of the other
experiments mentioned, rather than as experiments in and of themselves.  However,
the physics described in each could be tested directly by experiment.

\subsection{Lorentz invariant encodings}
The transmission of quantum states among parties in Earth orbit and on the
ground is complicated by the fact that the observers do not have an obvious
shared Lorentz frame against which to perform their measurements.  It is
possible to avoid this difficulty by encoding qubits into a noiseless
subsystem of several particles, in such a way that the subsystem is invariant
under Lorentz transformations~\cite{BT05,BRS003}.
Using the fact that photons of opposite helicities acquire opposite phases (Section~\ref{lorentz-polar}), an encoding of one logical qubit into two physical qubits makes the knowledge of  the relative orientation and velocities of the Lorentz frames redundant.
 By utilizing this
technique, parties can exchange entangled qubits without caring about the
relative Lorentz transformation which relates their respective reference
frames.

For the various experiments one can then decide whether the effect of the
Lorentz transformations are of interest, and separate it out when desired.
Thus one could isolate different effects with the same orbital scenario,
depending on interest.

The noiseless subsystem method is also applicable to the gravitomagnetic phase (Section~\ref{frame_dragging.sec}).

It is not clear what the effect of relative acceleration on these encoded
states would be.  One could presume this leads to higher order corrections. This question should be
investigated.

\subsection{Preparation contextuality}

Spekkens et~al.\ have proposed an alternative to the EPRB experiment
which, rather than directly testing the Bell inequalities, tests a
``preparation contextuality'' inequality, whose violation by quantum
mechanics manifests a sort of quantum entanglement~\cite{Spekkens-prep_context}.
The experimental setup is much simpler,
consisting of a state preparation by one party, followed by a single
measurement of a second party.  The intent is to send ``parity-oblivious''
quantum information about a bit string to a second party, who then queries the
information by performing a measurement on the received quantum state.

The major advantage of this technique is that it allows considerably greater
accuracy over the more conventional EPRB-type experiment.  One has achieved
98\% of the quantum bound on the preparation contextuality inequality~\cite{Spekkens-prep_context}.
Perhaps the improved accuracy will be important for testing unknown physics
in some of the proposed experiments.

A downside is that it is more difficult to see exactly what is being ruled
out by this test.  It is only relevant to some loopholes of the EPRB
experiment.  In particular, the condition of ``locality'' is replaced by a
parity obliviousness condition.  While the former is easy to determine by
simply checking for spacelike separation of the two measurement events, it is
not obvious how to physically determine if the transmission is truly parity
oblivious.

\section{Technology}
\label{sec.technology}

\subsection{Measuring the new effects}

To implement the experiments that will investigate the various effects described in the preceding sections, the necessary technologies must be engineered. In some cases, the necessary technology already exists sufficient to demonstrate the expected effects, and experiments could be carried out in the near term (granted appropriate funding). Indeed, proposals for such near-term space-based quantum missions are presently being evaluated by the CSA, the European Space Agency (ESA), and others. On the other hand, some experiments will require technological capabilities that are not currently available, and such missions will necessarily be long-term.

One notable challenge stems from the fact that the expected observations require precise knowledge of position, orientation, and/or timing in order to reach the maximal level predicted, or in some case to even be significantly evident beyond random statistical variation. The challenge facing space experiments is the difficulty of performing this reference frame alignment without invoking the quantum effects that the experiment itself aims to demonstrate (and thereby maintaining the validity of the experiment's results). The alignment of source and receivers for quantum signals should be based on an independent, common reference, e.g.\ surrounding fixed stars and other celestial objects.

Even with the precise reference frame alignment, the quantum effects that we expect to observe will, in most cases, remain very small. The challenge in this situation is, then, to convincingly distinguish between the quantum contributions and statistical noise within the experimental measurements. For experiments utilizing orbiting satellites, a potential avenue to aid this task is to utilize the inevitable variation in the satellite's orbit. Because the quantum effects of interest have dependence on the height and/or speed of the apparatus, with a suitably eccentric elliptical orbit, a pattern would emerge within the oscillations of the measurement data, correlating to the orbital position of the satellite.

\begin{figure}
\centering\includegraphics[width=0.5\textwidth]{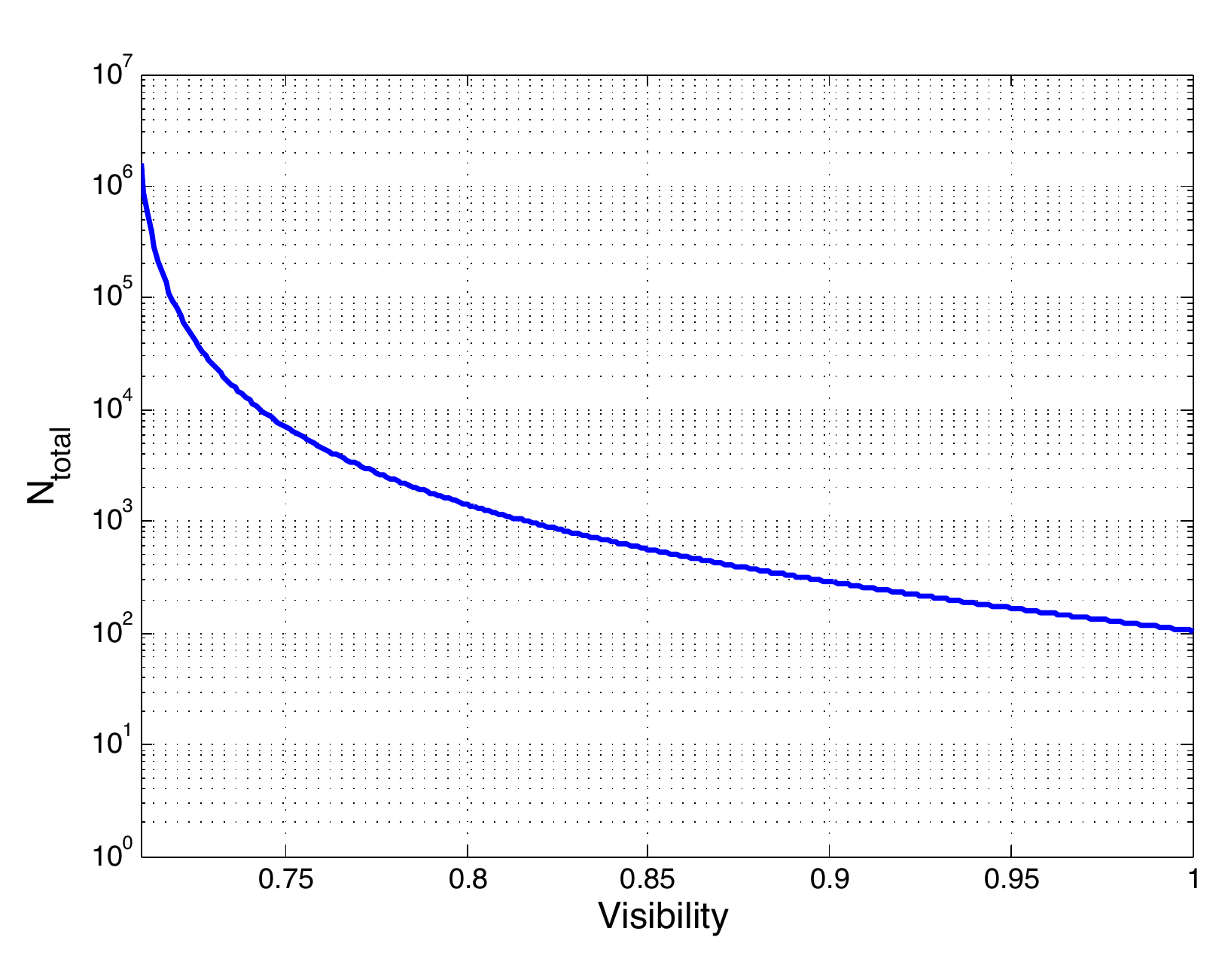}
\caption{The number of photon detections $N$ required in order to achieve a 3$\sigma$ violation of the Bell inequality, given an entanglement visibility $V$.}
\label{fig:requiredNforBell}
\end{figure}

With or without such tricks, a minimum number of measurements would be necessary to achieve statistical significance in most cases. In the case of Bell tests (Section~\ref{sec:belltest}), the number of measurements required to achieve a significant (at least $3\sigma$) violation of the Bell inequality depends on the fidelity of the quantum signals received. Utilizing the Poissonian statistical nature of photon detections, error propagation through the correlation functions that make up the Bell inequality leads to the conclusion that
\begin{equation}
 N > \frac{36\left(1-V^2/2\right)}{\left(\sqrt{2}V-1\right)^2}
\end{equation}
photons must be detected, given an experimental entanglement visibility $V$ (see Figure~\ref{fig:requiredNforBell}).

\subsection{Eliminating known sources of noise from the signals}

As in any experiment, there will be inevitable imperfections in the design and manufacture of components, their calibration, and their operation. Moreover, there also exist physical restrictions (for example, the diffraction limit) that bound the precision of measurements that can be made using practical apparatuses. These effects will add noise and uncertainty to the results obtained from any experimental apparatus deployed in a space platform. How do we then separate the ``regular'' sources of noise from any as yet unknown decoherence effect?

Such a discussion may require careful and thorough theoretical analysis encompassing all the known noise contributions, e.g.\ photon loss, erroneous detection of background light and thermally induced (dark) counts. All these effects introduce a predictable amount of noise, thus degrading the entanglement visibility. However, the measurement of significant deviation from the expected degradation would be indicative of new physics (assuming the models are as complete as possible).

\subsection{Proposed systems}

Proposals for future missions to test quantum effects in space are currently being considered by a number of space agencies worldwide. These missions will, if implemented, make substantial primary steps in observations of large-scale quantum effects. Space-QUEST (``QUantum Entanglement for Space ExperimenTs''), an initiative of the ESA and led by the Institute for Quantum Optics and Quantum Information in Vienna, proposes to place an entangled photon pair source on the International Space Station (ISS)~\cite{Armengol2008,Ursin_EPN_Space-QUEST}. It would possess two independently orientable telescopes of 10--15\,cm diameter which will transmit the entangled photons to ground receivers at locations separated by over 1000\,km\footnote{An alternate configuration has only one orientable telescope. One of the photons generated by the source would instead be detected at the ISS.}, allowing long-distance Bell tests, quantum key distribution, and other science experiments.

A notably simpler approach is being considered by the CSA. The QEYSSAT (``Quantum EncrYption and Science SATellite'') proposal, led by the Institute for Quantum Computing in Waterloo, reverses the direction of the quantum transmission, placing the complicated parts of the apparatus --- the photon source --- on the ground, and the simple part --- the photon receiver --- on a microsatellite in a noon-midnight LEO orbit. This approach experiences higher loss compared to Space-QUEST's downlink approach, but it also possesses several practical advantages in source interchangeability and ease of maintenance on the ground, as well as lower space, power, and classical processing requirements on the satellite. Despite its simplified nature, QEYSSAT will also allow testing of long distance entanglement (also over 1000\,km), and other scientific questions such as Ralph's gravity-induced entanglement decorrelation (Section~\ref{sec.Ralph}) and simultaneity paradoxes.

\section{Conclusion}
\label{sec.conclusion}

The advent of technologies which enable quantum communication in Space has
the potential to open a new chapter in the development of our understanding
of the physical world.  Tests of quantum phenomena are no longer bound to
laboratory settings, allowing us to explore the nature of the quantum
world at increasingly large length scales.
In doing so, there lies before us the potential to find new revelations
affecting the foundations of 
quantum field theory and general
relativity, largely unchanged for almost a century, that now form the pillars of modern physics.

We have discussed a wide variety of potential experiments to probe physics in
every known regime, from ordinary non-relativistic quantum mechanics to ideas
stemming from various attempts to reconcile gravitational physics with
quantum theory. 
For
most experiments we attempt to indicate its feasibly in terms of technology
required, and the magnitude of an effect which one might hope to measure.

We hope that this survey article
will stimulate others to further investigate some of these ideas,
seriously considering what possibilities for
experimental investigation of fundamental physics are, or can be, enabled by
these technologies.

\section{Author contributions}
All authors participated in the discussions about performing quantum science experiments in Space. Initiators of these discussions were T.J., D.R. and R.L. in response to a call by the Canadian Space Agency for identifying possible near-term and long-term tests of quantum entanglement science in Space.
The primary contact for each section are:
T.J. and D.R. for Section~\ref{timeframes.sec} (\nameref{timeframes.sec}),
T.J., D.R. and B.L.H. for Section~\ref{sec.entanglement} (\nameref{sec.entanglement}),
A.Kent for Section~\ref{macro_amp.sec} (\nameref{macro_amp.sec}),
J.M. for Section~\ref{bimetric.sec} (\nameref{bimetric.sec}),
D.T. and T.D. for Section~\ref{sec.entanglementSRGR} (\nameref{sec.entanglementSRGR}),
A.Kempf and D.T. for Section~\ref{sec.parallel_transport} (\nameref{sec.parallel_transport}),
E.M-M. for Section~\ref{Emm1} (\nameref{Emm1}),
R.M., N.C.M. and D.T. for Section~\ref{cow.sec} (\nameref{cow.sec}),
E.M-M. and R.M. for Section~\ref{unruh.sec} (\nameref{unruh.sec}),
X.M. and N.C.M. for Section~\ref{sec.Ralph} (\nameref{sec.Ralph}),
E.M-M. for Section~\ref{EMM2} (\nameref{EMM2}),
D.R. for Section~\ref{diffusion.sec} (\nameref{diffusion.sec}),
G.A-C. for Sections~\ref{noncommutativity.sec} (\nameref{noncommutativity.sec})
and~\ref{DSR.sec} (\nameref{DSR.sec}),
T.J., B.L.H. and C.S. for Section~\ref{sec.qcommunication} (\nameref{sec.qcommunication}),
A.Kent for Section~\ref{tagging.sec} (\nameref{tagging.sec}),
B.L.H., T.J. and D.R. for Section~\ref{sec.spekkens} (\nameref{sec.spekkens}),
and B.L.H. and T.J. for Section~\ref{sec.technology} (\nameref{sec.technology}).

\section{Acknowledgements}
The authors acknowledge support for this study by the Canadian Space Agency,
CIFAR, NSERC, Industry Canada, as well as the Perimeter Institute for
Theoretical Physics for hosting
this series of discussions. 
Research at the Perimeter Institute is supported by the
Government of Canada through Industry Canada and by the Province of Ontario through the Ministry of Research
and Innovation.
The work of D.R. has been supported in part
by the Defense Advanced
Research Projects Agency as part of the Quantum Entanglement Science
and Technology program under grant N66001-09-1-2025,
and by a grant from the Foundational
Questions Institute (FQXi) FQXi-RFP3-1018.
A.K. was partially supported by a Leverhulme Research Fellowship, and a grant
from the John Templeton Foundation.
JWM thanks the John Templeton Foundation for the generous support of his research.

We thank the following people for valuable discussions: Rob Spekkens, Norbert
L\"utkenhaus, Ian D'Souza (COM DEV), Danya Hudson (COM DEV), Ralph Girard (CSA), Chris
Erven, Catherine Holloway, Evan Meyer-Scott, and Daniel Gottesmann.

\end{document}